\documentclass[journal,twoside,web]{ieeecolor}
\usepackage{generic}
\usepackage{subfig}
\usepackage{multirow}
\usepackage{cite}
\usepackage{amsmath,amssymb,amsfonts}
\usepackage{algorithmic}
\usepackage{graphicx}
\usepackage{algorithm,algorithmic}
\usepackage{hyperref}
\hypersetup{hidelinks}
\usepackage{textcomp}
\def\BibTeX{{\rm B\kern-.05em{\sc i\kern-.025em b}\kern-.08em
    T\kern-.1667em\lower.7ex\hbox{E}\kern-.125emX}}
\providecommand{\refname}{References}


\begin{document}
\title{i-Socket: Design, Development, and Pilot Evaluation of an Individualized Multi-Material, Multi-Thickness Transtibial Prosthetic Socket}
\author{Noor~Alhuda~Ameen, Omid~Arfaie, and~Ramazan~Unal
\thanks{This study is funded by the Scientific and Technological Research Council of Turkey (TUBITAK) under the project number 123M351. }
\thanks{The authors are with the Human-Centered Design Laboratory, Department of Mechanical Engineering, Ozyegin University, Istanbul, Turkey (Corresponding author: e-mail: ramazan.unal@ozyegin.edu.tr).}}

\maketitle

\begin{abstract}
The prosthetic socket is an essential part in ensuring comfort and stability for the overall prosthesis system. This study proposes a multi-material/thickness individualized transtibial prosthetic socket that focuses on providing comfort. This study aims to identify the proper material and thickness to protect Calf areas and reduce the pressure around bone areas. First, the socket is divided into four parts depending on the pressure-sensitive/tolerant regions. After identifying the thickness range for each concerned area, the thickness and material are selected based on the Pressure-Pain Threshold (PPT) test, and the finalized design is then prototyped. The prototyped individualized socket (i-Socket) is 22\% lighter than the participant’s own socket. Results of the pilot experiments with an amputee participant showed that the pressure inside the socket decreased by 45\% and 31\% for the Tibia and Fibula regions, respectively. Additionally, the self-selected CoM velocity for the walking experiment is increased by 15\% compared to the similar studies in the literature. Regarding the kinematic results, symmetry in the knee and ankle joints increased by 65\% and 2\% when using the i-Socket compared to the results with the participant's own socket. 
\end{abstract}


\section{Introduction}
\IEEEPARstart{I}{t} is documented that the rejection or abandonment rate of prostheses is estimated to be between 24\% and 70\% \cite{who2017}. The reasons behind rejecting or abandoning the prostheses are discomfort while using the prostheses, the quality of the prostheses, the prosthetic appearance, and the high cost of the prosthesis \cite{varsavas2022review}. The rationale behind this is several interconnected factors, one of which is high-pressure distribution, and a heavy socket that induces discomfort, instability, pain at certain places in the body, insufficient socket fitting, skin damage, patient fatigue, and weakening of the socket’s structural integrity. 
The socket is the interface between the user and prosthesis, where control and stability are provided from the socket to the entire prosthesis \cite{noll2017physically}. Therefore, a good interface between the residual limb and socket is essential for a successful prosthetic fitting and rehabilitation of a transtibial amputee \cite{xiaohong2004dynamic}. Consequently, a poor socket fit disrupts this interface, reducing the amputee's mobility. The lack of mobility contributes to muscle atrophy, and the physical stress of loading causes skin issues and back pain \cite{price2019design}. However, a poor socket fit can increase these problems or directly trigger these impairments \cite{price2019design}. Similarly, if high interface pressure between the residual limb and prosthetic socket occurs, participants will experience instability, discomfort during daily activities, and pain or skin damage \cite{dumbleton2009dynamic, ibarra2020interface, paterno2018sockets}. Moreover, heavy socket contributes to several negative outcomes, mostly unease, patient fatigue, and instability due to the possibility of exerting excessive stress on the suspension system \cite{gariboldi2022structural}. As a result of this, abandoning the prosthetic occurs, leading to decreased locomotion \cite{mak2001state}.  

A commonly used approach for reducing pressure around the residual limb by adding extra contact and hollowed area for tolerated pressure and sensitive region, and it’s called the Patella Tendon Bearing (PTB) prosthetic socket for transtibial amputees \cite{yiugiter2002comparison}. Moreover, a Total  Surface Bearing (TSB) prosthetic socket is also used, and it evenly distributes the pressure across the residual limb by keeping the residual limb in full contact and applying gentle pressure to the Calf area. The TSB socket shows better results than the PTB socket in terms of suspension, pressure distribution, and socket mass. However, the PTB and TSB socket masses are around 1.11 kg and 1.04 kg, respectively \cite{yiugiter2002comparison}. Nevertheless, a participant uses PTB and TSB sockets for 1.5 years, both of which fail to relieve pressure around the sensitive region \cite{bagheripour2022design}. 
An additional approach using reinforcement, like carbon fiber, fiberglass, and cement, on a 3D-printed Polylactic Acid (PLA) socket, enhances structural strength and reduces socket failure under load. The approach aims to improve the mechanical performance of prosthetic sockets without complicating the design \cite{ramlee2024investigation}. Although cement coating results in the highest strength, further research is needed to evaluate long-term strength \cite{ramlee2024investigation}. However, the study does not investigate comfort or pressure distribution, which limits addressing its effect on residual limb comfort and pressure mapping.
Another approach uses Selective Laser Sintering (SLS) and a Finite Element Analysis (FEA) framework to build the prosthetic socket made from Duraform polyamide (PA) with localized reductions in wall thickness to evaluate structural performance. The design is validated by comparing the deformation and failure loads, and the results show that the predicted failure loads match within 3\% and the deformation at the bottom of the socket is within 25\% of the FEM predictions under sub-fault loads \cite{faustini2006experimental}. The pressure distribution applied in the model is not patient-specific, and it is not feasible to obtain accurate values for each socket design \cite{faustini2006experimental}. Nonetheless, the evaluation does not include a pressure map distribution for the socket design, which is for assessing how these changes in wall thickness affect pressure distribution. 
An ongoing alternative solution for a transtibial prosthetic socket involves designing a multi-material 3D-printed socket. The study aims to design comfortable and low-cost prosthetic sockets by targeting the load and off-load zones. The proposed idea is that load zones are covered with PLA material and soft material, like rubber material, that covers the off-load zones. In addition, different infills are tested to achieve a suitable balance of resistance and mass of the socket \cite{comotti2015multi}. Although the concept of surface interaction between the socket and lower limb through load and off-load is valid, time is required to confirm the materials' safety and reliability for daily activities \cite{comotti2015multi}. Further study focuses on designing Variable Impedance Prosthetic (VIPr) sockets using computer-aided design and manufacturing using TangoBlackPlus (rigid material) and VeroWhitePlus (flexible material). Studying the depth of bones and connecting it with the Young's modulus of the materials will help in material selection \cite{sengeh2013variable}. Moreover, when the bone is far from the skin, the material is chosen to be stiff, and when the bone is close to the skin, the material should be flexible. When the participant's speed is compared while wearing the VIPr socket and the participant's socket, it appears that the participant's speed increases by 16\% while wearing the VIPr socket \cite{sengeh2013variable}. In addition, the mass of the VIPr is almost three times heavier than the participant's own socket (Carbon fiber material), and this is because the socket wall is thicker to achieve structural integrity and due to material properties limitations \cite{sengeh2013variable}.
This study is aimed at individualizing transtibial prosthetic sockets based on designing a multi-material/thickness socket according to the participant’s PPT in sensitive areas. The PPT test is employed to determine the material and thickness of each region for improving comfort with pressure reduction around the pressure-sensitive/tolerant areas. The areas that are considered in the study are the Tibia, Fibula, and Calf areas. Multiple wall thicknesses are selected, based on a design study conducted using existing pressure values. After this design study, a PPT test is conducted to finalize material and thickness selection for each area, depending on the participant’s test results. To ensure the socket is fitted with the participant's residual limb, the residual limb scanning is conducted in a careful way. Also, to avoid a negative impact on stability, the prosthetic socket mass, approximately, remains the same as the participant’s own socket \cite{gariboldi2022structural}. A pressure map distribution for the concerned area is obtained during a static test conducted using a universal compression machine. The results for the specified areas show that the pressure in these areas is reduced compared to the pressure results of existing sockets in the literature. Additionally, the functional evaluations by a transtibial amputee are performed to investigate the pressure distribution inside the socket and the kinematic performance during walking at self-selected speed. The results show that the pressure and kinematic metrics are improved while using the individualized socket (i-Socket) compared to the results with the participant's own socket (POS).
\section{Methods}
In this section of the study, the methodology for designing and identifying the thickness and material selection procedure is presented.
\subsection{Conceptual design}
\begin{figure}[H]
    \centering
    \includegraphics[width=\columnwidth]{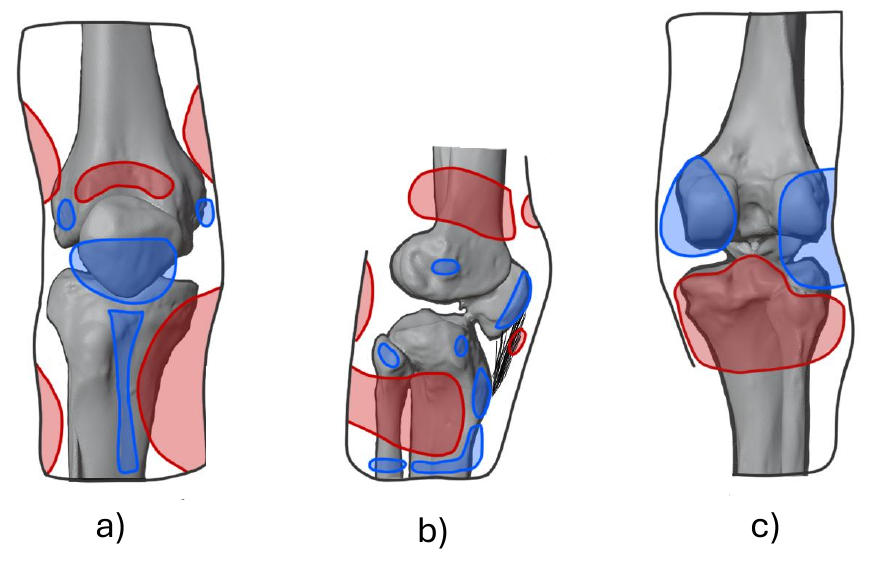}
    \caption{This diagram shows the pressure-tolerant and pressure\-sensitive areas \cite{physiopedia_lower_limb_sockets}. Anterior view (a), Lateral view (b), Posterior view (c). Pressure-sensitive areas are highlighted in blue, and Pressure-tolerant areas are highlighted in red.}
    \label{critical region}
\end{figure}

Designing a comfortable socket depends on distributing the pressure in the right areas. The areas that tolerate pressure need a rigid material to avoid high pressure, and the pressure-sensitive areas need a flexible material. To determine areas that need protection or require flexibility, a PPT test should be performed on the participant, but initially, the soft tissue and bone areas are assumed, depending on pressure-tolerant and pressure-sensitive regions, as shown in Fig.\ref{critical region}. The upcoming subsections provide detailed information about all designs and analyses. 

\begin{figure}[H]
    \centering
    \subfloat[]{
        \includegraphics[width=0.46\columnwidth]{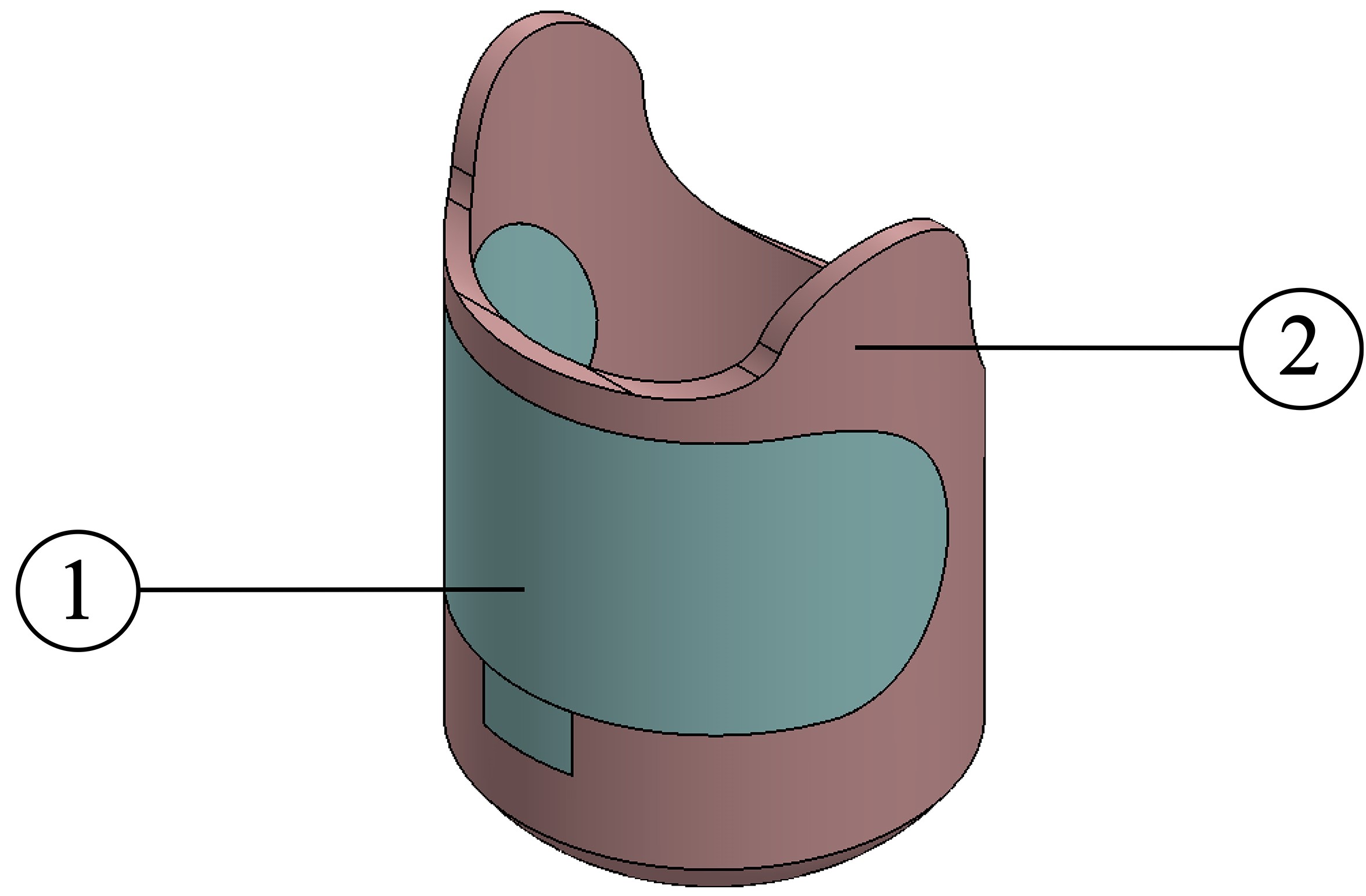}
        \label{Design1}
    }
    \subfloat[]{
        \includegraphics[width=0.46\columnwidth]{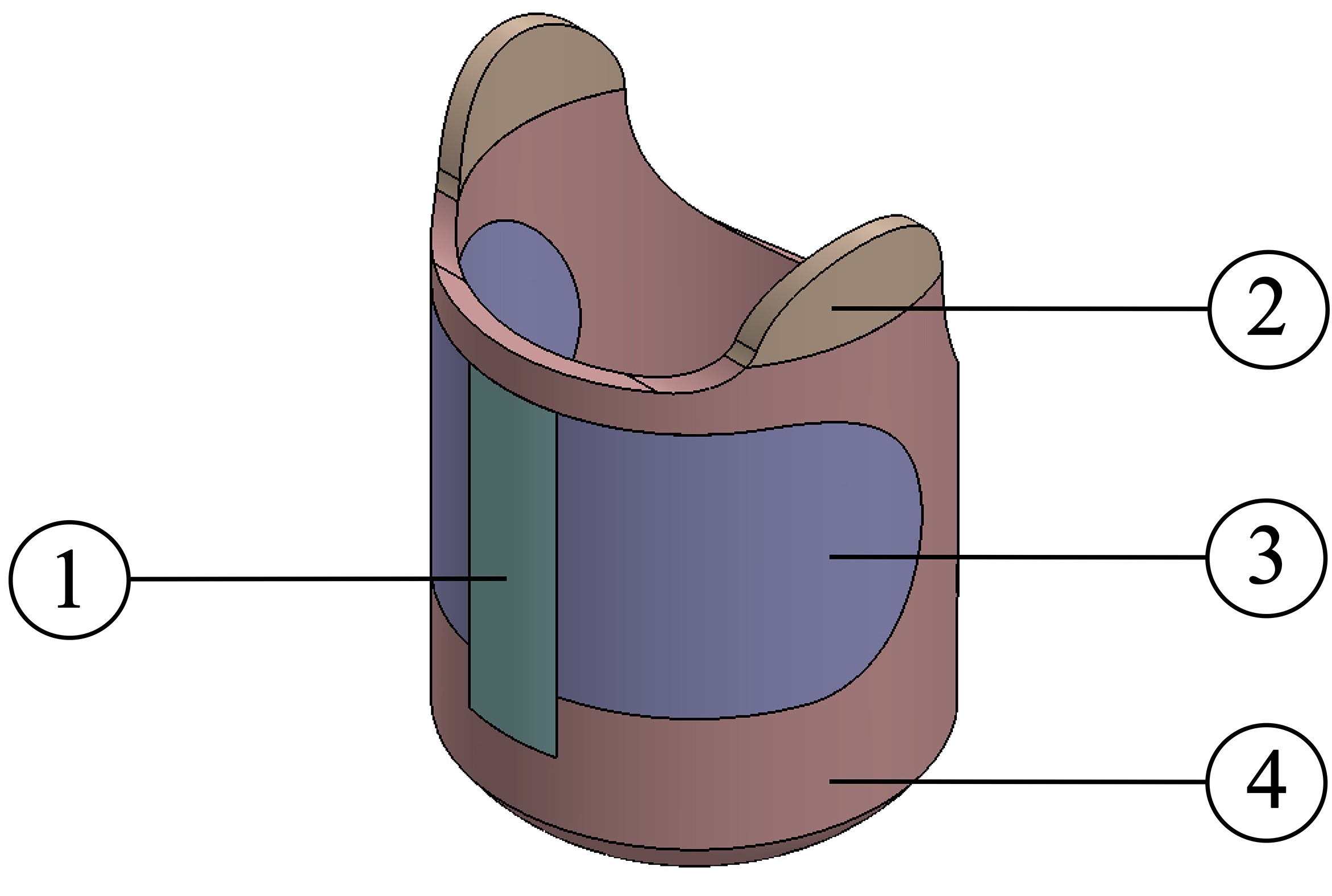}
        \label{Design2}
    }
    \caption{CAD model of Design 1 (a): Pressure-sensitive areas (1) and pressure-tolerant areas (2). CAD model of Design 2 (b): Pressure-sensitive areas (1\&3) and pressure-tolerant areas (2\&4).}
    \label{fig:design_comparison}
\end{figure}

At first, hypothetical models are built to observe the stress acting on the socket during static analysis with a diameter of 0.0962 m and a height of 0.16 m. The first socket design addresses pressure-tolerant areas by using rigid materials, with carbon fiber and Kevlar to be evaluated. Pressure-sensitive areas are accommodated through the placement of a soft and flexible material, and in this design, Thermoplastic Polyurethane (TPU) material is considered. The material arrangement for Design 1 is presented in Fig. \ref {Design1}. As an initial assumption before determining the thickness, the thickness of Design 1 is considered to be 5 mm.

\begin{figure}[H]
    \centering
    \includegraphics[width=0.6\columnwidth]{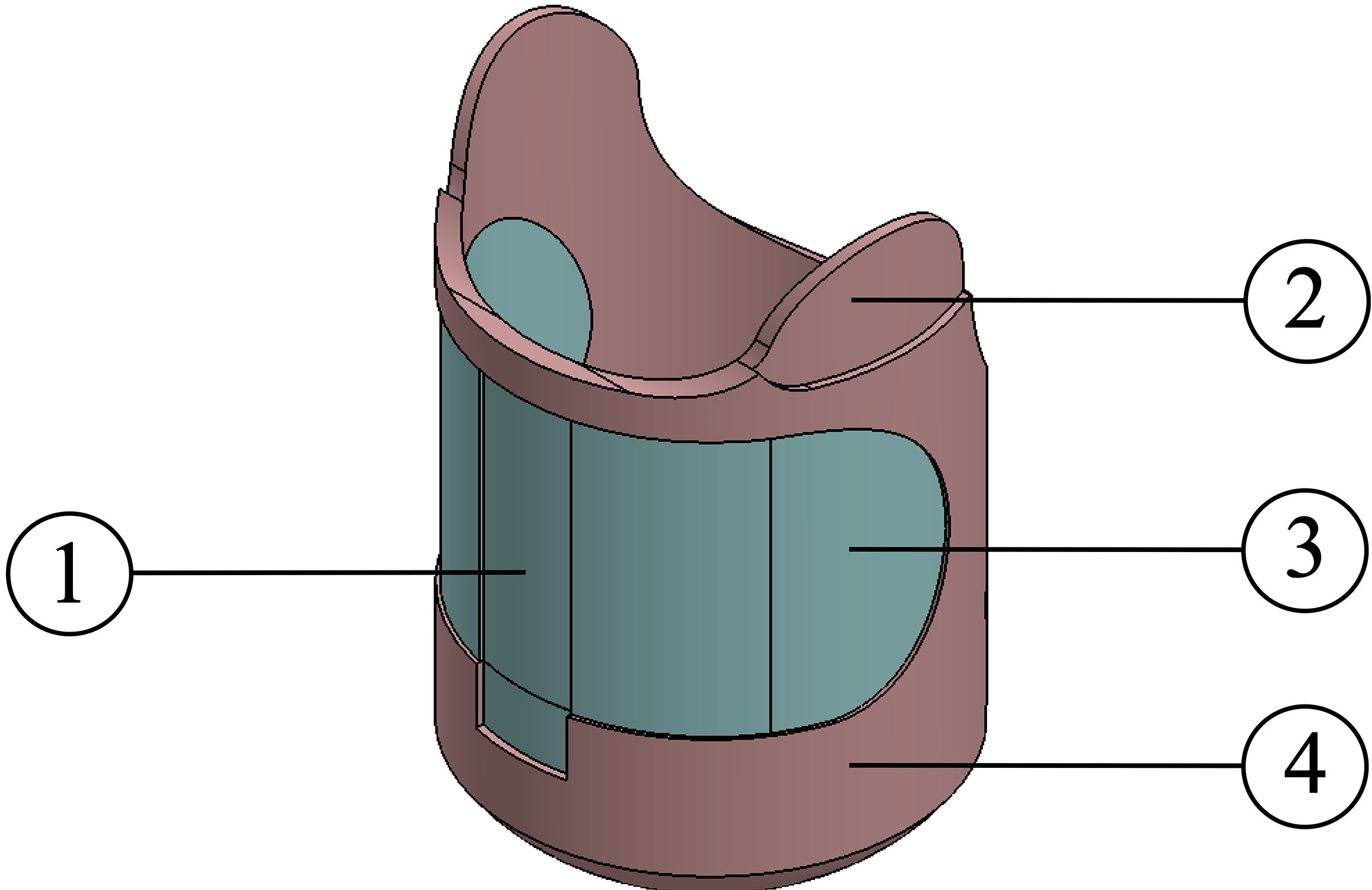}
    \caption{CAD model of Design 3: Pressure-sensitive areas (1 \& 3), and Pressure-tolerant areas (2 \& 4).}
    \label{Design3}
\end{figure}

Design 2 is divided into four sections, as shown in Fig.\ref{Design2}. The socket contains four materials, with Kevlar and Carbon fibers used for pressure-tolerant areas, and PLA and TPU for pressure-sensitive areas. Furthermore, as in design 1, all areas have the same thickness of 5 mm.
For achieving a high material rigidity or flexibility, the thickness of each area can be changed according to the output needed, and in addition, changing the thickness helps in reducing the socket's overall mass. Design 3 integrates the material layout distribution of design 2 with the material selection of design 1, but modulating the thickness of each region to achieve the desired rigidity and flexibility, as illustrated in Fig.\ref{Design3}. The thicknesses are initially assumed, with pressure-tolerant areas at 5.00 and 3.00 mm, and are tested with both Carbon fiber and Kevlar, while pressure-sensitive areas have thicknesses of 4.00 and 3.00 mm, manufactured from TPU.

\begin{figure}[H]
    \centering
    \includegraphics[width=\linewidth]{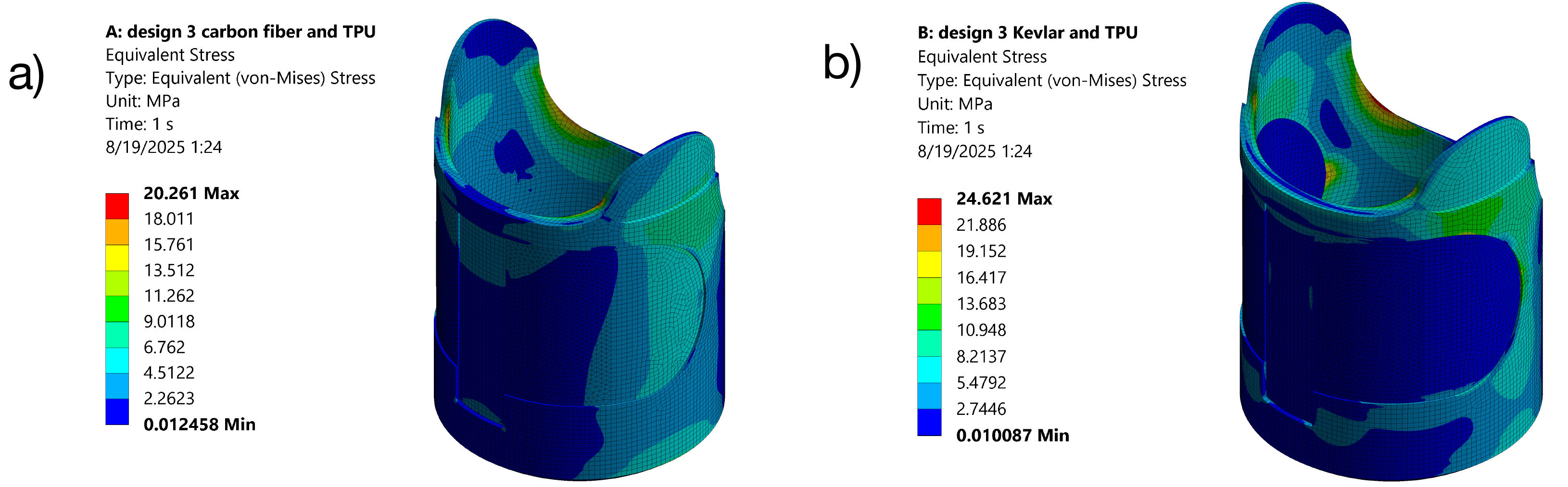}
    \caption{Design 3 structural analysis, a) Stress analysis using Carbon fiber for pressure-tolerant areas, b) Stress analysis using Kevlar for pressure-tolerant areas.}
    \label{Design3 structural analysis}
\end{figure}

For all designs, a stress analysis is conducted according to the forces and pressures acting between the socket and residual limb during the stand phase \cite{sengeh2013variable, novacheck1998biomechanics}. The stress analysis for design 3 provides critical insight from material and design perspectives. According to the results and from a material perspective, carbon fiber transmits stress to adjacent regions, while Kevlar reduces stress in medium and flexible areas, as presented in Fig.\ref{Design3 structural analysis}. 

Additionally, from the design perspective, different thicknesses don’t necessarily decrease maximum stress; they help redistribute stress across regions and control rigidity and flexibility. Based on stress analysis, design 3 reduces localized peaks in sensitive areas compared to designs 1 and 2, and with multiple thicknesses, stress is distributed evenly across the socket. To address all the study's requirements, design 3 is considered the final design due to its ability to redistribute stress and control rigidity and flexibility.

\subsection{Thickness range determination} 
To specify the range of thicknesses for the areas concerned in the final design, which satisfies the study's requirements, parametric research is conducted based on the PPT of healthy and amputee participants in the literature. Since amputee PPT test data are not provided for all regions, healthy subjects' PPT test values are used for the thickness range determination of the Tibia region.

To identify the thickness in the Calf area, the PPT value of 0.438 MPa with a sensor area of 100 $mm^2$ is considered as the constraint for this region \cite{novacheck1998biomechanics}. Using a parametric analysis, the desired thickness for this area is identified. The inner pressure in this area is considered as the standing load on the socket, approximately equal to 0.100 MPa \cite{convery1999socket}. For the Calf area, three thicknesses are considered as 5.00 mm, 7.50 mm, and 10.00 mm. For thicknesses below 5.00 mm, the analysis fails, and a thickness of more than 10.00 mm will increase the socket’s mass. The study results are presented in Table \ref{Table 1}. The results show that thicknesses of 7.50 mm and 10.00 mm do not exceed the constraint, and the pressure values are almost identical for these two thicknesses. Hence, to achieve a low-weight prosthetic socket, a thickness of 7.50 mm is considered.

\begin{table}[H]
    \centering
    \caption{The results of the pressure on the Calf area for different thicknesses}
    \label{Table 1}
    \renewcommand{\arraystretch}{1.3} 
    \begin{tabular}{c c c} 
        \hline
        \textbf{Thickness [mm]} & \textbf{Stress constraint [MPa]}  & \textbf{Study Result [MPa]}  \\
        \hline
        5.00 & \multirow{3}{*}{Study Result $<$ 0.438} & 0.505 \\ 
        7.50 & & 0.353 \\ 
        10.0 & & 0.373 \\ 
        \hline
    \end{tabular}
\end{table}

For the Fibula area of the residual limb, the PPT test value is reported as 0.490 MPa \cite{lee2005regional}, but the utilized sensor area around this region is not specified. Based on the findings in the literature \cite{convery1999socket}, the applied pressure on the Fibula area during standing is reported as 0.03 MPa. Since the Fibula region is considered a pressure-sensitive area, to achieve both strength and lightness in the socket, the selected thickness for the parametric study for this area is considered to be between 3.00 mm and 6.00 mm. According to the results presented in Table \ref{Table 2}, all thicknesses are considered in the range since none of the thicknesses exceed the stress constraint.

\begin{table}[H]
    \centering
    \caption{Four considered thicknesses in the parametric study and their results for the Fibula area}
    \label{Table 2}
    \renewcommand{\arraystretch}{1.3}
    
    \begin{tabular}{c c c} 
        \hline
        \textbf{Thickness [mm]} & \textbf{Stress constraint [MPa]}  & \textbf{Study Result [MPa]}  \\
        \hline
        3.00 & \multirow{4}{*}{Study Result $<$ 0.490} & 0.278 \\ 
        4.00 & & 0.257 \\ 
        5.00 & & 0.236 \\ 
        6.00 & & 0.237 \\ 
        \hline
    \end{tabular}
\end{table}

The contact pressure around the Tibia area is nearly 0.01 MPa during the standing task \cite{convery1999socket}. Accordingly, to identify the thickness in this region, the PPT test value of the Tibia region for healthy participants is reported as 0.454 MPa, and the sensor contact area is 100 mm² \cite{rolke2005deep}. The thickness range considered in this study is from 3.00 mm to 6.00 mm, as shown in Table \ref{Table 3}, which includes all the thicknesses in the range from the parametric study of the Fibula area.

\begin{table}[H]
    \centering
    \caption{Four considered thicknesses in the parametric study and their results for the Tibia area}
    \label{Table 3}
    \renewcommand{\arraystretch}{1.3} 
    \begin{tabular}{c c c} 
        \hline
        \textbf{Thickness [mm]} & \textbf{Stress constraint [MPa]} & \textbf{Study Result [MPa]} \\
        \hline
        3.00 & \multirow{4}{*}{Study Result $<$ 0.454} & 0.111 \\
        4.00 & & 0.103 \\
        5.00 & & 0.098 \\
        6.00 & & 0.096 \\
        \hline
    \end{tabular}
\end{table}

Based on the study results, 5.00 mm and 6.00 mm thickness results are similar, as presented in Table \ref{Table 1} and Table \ref{Table 2}. Due to this, 5.50 mm is included in the range. The final thickness ranges for the PPT test are considered to be 3.00 mm, 4.00 mm, 5.50 mm, and 7.50 mm.

\subsection{PPT Experiments}
\begin{figure}[H]
    \centering
    \subfloat[]{
        \includegraphics[width=0.45\columnwidth]{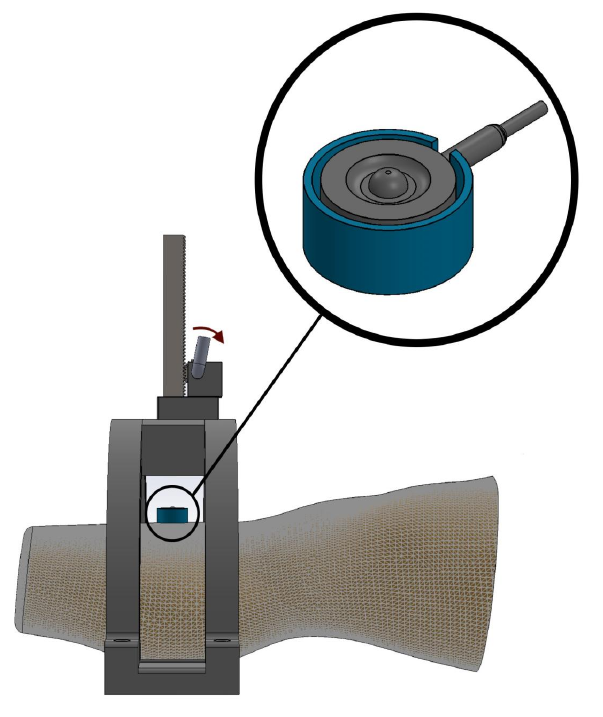}
        \label{SensorSETUP}
    }
    \subfloat[]{
        \includegraphics[width=0.49\columnwidth]{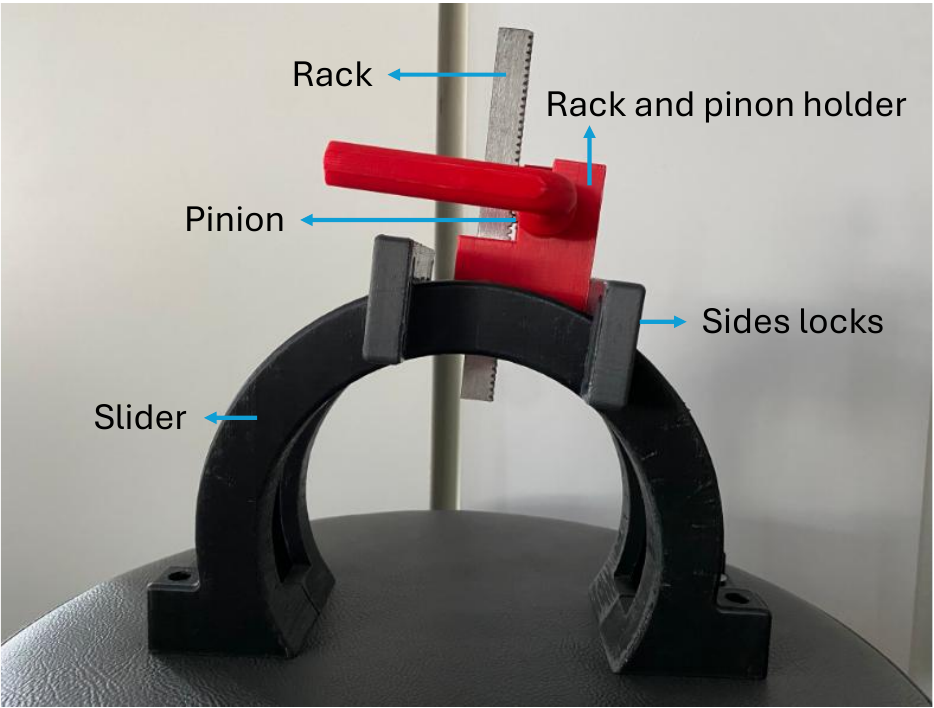}
        \label{printedsetup}
    }
    \caption{PPT setup: - from tough PLA (a), 3D model for the setup, residual limb, specimen, load cell, and force applier (b).}
    \label{PPT-all}
\end{figure}

Four different materials with various thicknesses are tested in the PPT test to finalize the participant's socket materials and thicknesses for the concerned areas. To build upon these principles, the following discussion is a review of existing PPT test methods for healthy and transtibial participants.
An existing study uses a pneumatic tourniquet connected to the computer-controlled air compressor, which is wrapped around the area to be tested \cite{polianskis2001computer}. In addition, the compressor is connected to the electro-pneumatic converter for converting the voltage into proportional output pressure, controlled via a data acquisition card. A software program is written in LabVIEW 5 for pressure control. The pain intensity is recorded using an electronic VAS. The compression cycles are carried out at 5-minute intervals, and the compression rate is set to 1.2 kPa/s. Another cycle has three compression rates: 0.25, 0.5, and 0.1 kPa/s. Another approach presented in the literature, a comparison for measuring PPT is conducted between two spring-loaded PTM (100-1000 kPa and 200-2000 kPa) and a pressure pain algometer (20-2000 kPa) [18]. Three regions are tested, including nail bed (thumb/ big toe), bone (Processus styloideus, medial malleolus), and muscle (Thenar eminence, abductor hallucis). The two PTM devices show close results, while the algometer’s pressure results for the nail bed and bone region are less \cite{rolke2005deep}. The following study measures the PPT for 11 areas on the residual limb by applying pressure to the skin through a circular material connected to a load cell \cite{lee2005regional}. For the participant, a 15-minute rest with the knee extended is applied before the test, and a 10-minute rest is provided between tests. The materials used in this test are pelite (mainly used for liners) and polypropylene (used for manufacturing prosthetic sockets). The load rate is controlled manually at 4 N/s  and a maximum force of 200 N \cite{lee2005regional}.
According to the information gathered from the existing setups \cite{polianskis2001computer,rolke2005deep,lee2005regional}, a setup for measuring the PPT is designed, as shown in Fig.\ref{printedsetup}. The setup has a semi-circle shape with a slider that moves left and right to target all the regions that need to be tested. Moreover, the slider ensures that the force applied is perpendicular to the skin. The force is applied using a rack and pinion mechanism, with a module of 1.0, a pinion with 25  teeth. The setup is 3D printed using tough PLA material. 

A specimen of TPU, Tough PLA, Kevlar, and Carbon fiber with the thicknesses found from the parametric study is printed, and each specimen is tested on the three areas (Tibia, Fibula, and Calf). Before starting the test, the participant rests for 15 minutes with their residual limb extended on the table and rests for 10 minutes between each test \cite{lee2005regional}. Furthermore, a linear force (maximum force 200 N) is applied on the specimen using the rack and pinion setup. When the gear’s handle rotates, the rack moves downward and generates an axial force. For measuring the applied force, a load cell (LMB-A, 500 N to 2 kN,  Kyowa, Japan) is placed at the top of the specimen, as shown in Fig.\ref{SensorSETUP}, and the load cell is connected to the Arduino Mega2560 board (Arduino, Somerville, MA, USA) using the HX711 amplifier. In addition, for data recording and calibration, the Basic Custom Arduino Library for HX711 on MATLAB is utilized. The entire setup is demonstrated in Fig.\ref{printedsetup}. The PPT values are determined when the participant starts to feel pain, and based on the results, the socket material of each area is selected.
The experiments of the i-Socket have been approved by the Clinical Research Ethics Committee of Istanbul Medipol University under document number E-10840098-772.02-7318. All of the participants in the functional evaluations have confirmed their participation by signing the informed consent form. A PPT test is conducted on six healthy participants using the PPT setup shown in Fig.\ref{PPT-all}. As mentioned in the previous section, the samples include the thickness range found from the thickness study. Furthermore, a 5 to 10-minute break is given between each sample. Each sample is tested twice to validate the PPT data, and the outcome is positive. The weak and strong areas differ from one person to another. While some of the participants endure the pressure with flexible materials, others endure it with rigid materials. Thicknesses of 1.00 mm and 2.00 mm are excluded due to immediate pain felt by the participants. The age range of healthy participants is between 19 and 22. The difference between the actual applied forces and the measured ones is considered as the error value of the device, which varies between 0.6\%-1.1\% for different participants.

\begin{figure}[H]
    \centering
    \includegraphics[width=\linewidth]{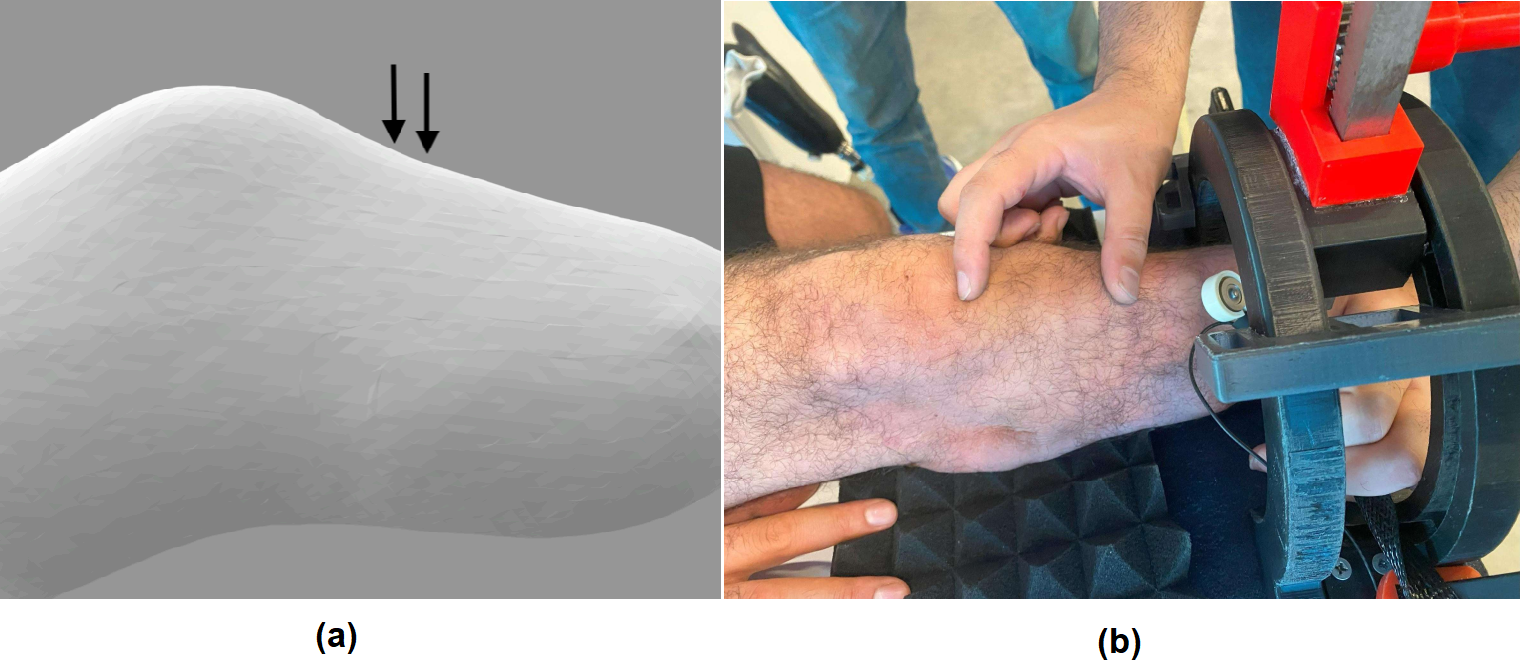}
    \caption{Participant pain area located at the top of the Tibia.}
    \label{painarea}
\end{figure}

Another PPT test is conducted for a single amputee participant on the Tibia, Fibula, and Calf regions. According to the results in Table \ref{Table 4} for this participant, the Tibia region cannot endure the pressure. Additionally, the participant is already experiencing pain in the area shown in Fig.\ref{painarea}. In daily activity, the participant does not encounter any pain in the Fibula and Calf areas, except that if the socket is placed incorrectly, the pain appears either on the same day or when wearing the socket again the next day. The 32-year-old participant has a body mass of 60 kg.

\begin{table*}[t]
    \centering
    \caption{The thickness of the socket for different materials of the Tibia, Fibula, and Calf regions}
    \label{Table 4}
    \renewcommand{\arraystretch}{1.3}

    \begin{tabular}{c l c c c c c c c c}
        \hline
        & & \multicolumn{2}{c}{\textbf{TPU}} 
        & \multicolumn{2}{c}{\textbf{Tough PLA}} 
        & \multicolumn{2}{c}{\textbf{Kevlar}} 
        & \multicolumn{2}{c}{\textbf{Carbon fiber}} \\
        \cline{3-10} 
        
        & \textbf{Thickness/Area} & \textbf{3.00} & \textbf{4.00} 
        & \textbf{3.00} & \textbf{4.00} 
        & \textbf{5.50} & \textbf{7.50} 
        & \textbf{5.50} & \textbf{7.50} \\
        \hline
        \multirow{3}{*}{\textbf{PPT [MPa]}} & \textbf{Tibia} & 0.152 & 0.229 & 0.060 & 0.067 & 0.099 & 0.060 & 0.050 & 0.038 \\
        & \textbf{Fibula}      & 0.188 & 0.183 & 0.188 & 0.272 & 0.297 & 0.183 & 0.319 & 0.228 \\
        & \textbf{Calf}        & -     & -     & -     & -     & 0.290 & 0.314 & -     & -     \\
        \hline
    \end{tabular}
\end{table*}

Regarding the PPT results, the final materials and thicknesses for the socket are presented in Table \ref{Table 5}. The PPT setup error percentage in the participant’s PPT test is approximately 1.1\%.

\begin{table*}[t]
    \centering
    \caption{Material and thickness selection for different socket regions}
    \label{Table 5}
    \renewcommand{\arraystretch}{1.3}
    
    \begin{tabular}{l l c c c l} 
        \hline
        \textbf{Participants} & \textbf{Properties} & \textbf{Tibia area} & \textbf{Fibula area} & \textbf{Calf Area} & \textbf{Rest of the socket} \\
        \hline
        \multirow{2}{*}{Amputee Participant} 
        & \textbf{Thicknesses [mm]} & 4.00 & 5.50 & 7.50 & 7.50 \\
        \cline{2-6} 
        & \textbf{Materials} & TPU & Carbon fiber & Kevlar & Tough PLA \\
        \hline
    \end{tabular}
\end{table*}

\subsection{Prosthetic Socket Model and Stance Phase Analysis}

\begin{figure}[H]
    \centering
    \includegraphics[width=0.75\columnwidth]{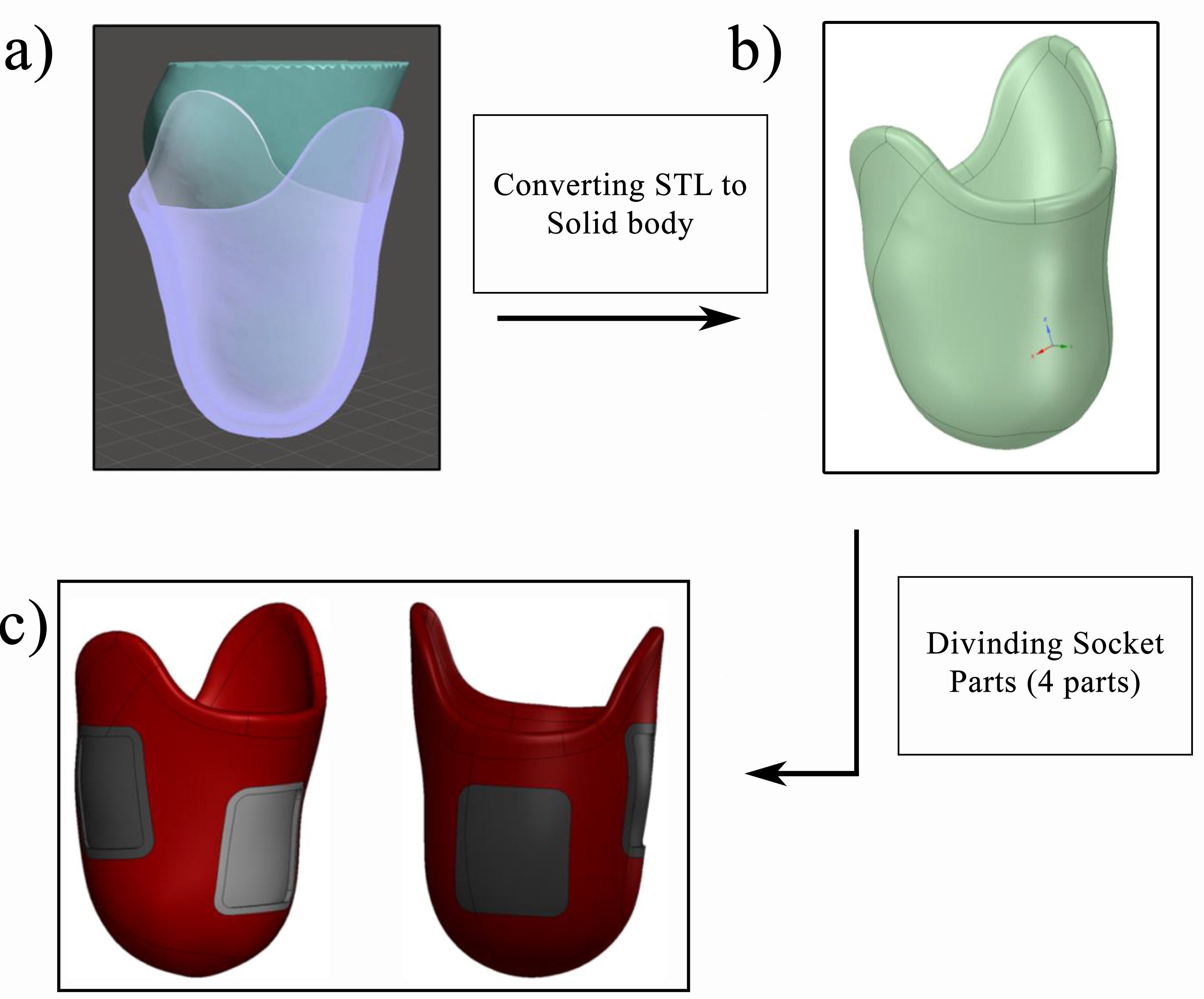}
    \caption{Extracted mesh data and socket 3D model (a), conversion of the socket design from mesh data to a solid part (b), composite parts separation (c).}
    \label{designstep}
\end{figure}

The steps of the CAD modeling for the socket are presented in Fig.\ref{designstep}. First, the residual limb is scanned, then the mesh data is extracted, and the socket design is done. Then, the socket is converted into a solid body, and its composite components are separated using CAD modeling software. In this final step, the thickness of each composite component determined through the PPT test is considered.
For obtaining the critical areas on the prosthetic socket, a transient structural analysis of the stance phase with a duration of 0.6s is conducted using material properties in Table \ref{Table 6}, the pressure values in Table \ref{Table 7}, and ground forces are calculated based on the participant’s mass and the knee flexion and extension, as shown in Table \ref{Table 8}. 

\begin{table}[H]
    \centering
    \caption{Materials properties that are used in this design.}
    \label{Table 6}
    \renewcommand{\arraystretch}{1.3} 
    \begin{tabular}{c c c c} 
        \hline
        \textbf{Material} & \textbf{Density} & \textbf{Young's modulus} & \textbf{Poisson} \\
         & \textbf{[kg/m$^3$]} & \textbf{[MPa]} & \textbf{ratio} \\
        \hline
        Carbon fiber & 1200 & 2600  & 0.3  \\
        TPU \cite{lee2019evaluation}     & 1450 & 2410  & 0.38 \\
        PLA          & 1250 & 3986  & 0.33 \\
        Kevlar \cite{markforged_composites_datasheet}  & 1200 & 27000 & 0.37 \\
        \hline
    \end{tabular}
\end{table}

\begin{table}[H]
\centering
\caption{Pressure values around the socket inner surface during the stance phase \cite{dumbleton2009dynamic}.}
\label{Table 7}
\renewcommand{\arraystretch}{1.3}
\begin{tabular}{l c c c}
\hline
\textbf{Area} & \textbf{Heel strike [MPa]} & \textbf{Mid-stance [MPa]} & \textbf{Push-off [MPa]} \\
\hline
Anterior  & 0.0574 & 0.0543 & 0.0754 \\
Medial    & 0.0698 & 0.0712 & 0.0836 \\
Lateral   & 0.0677 & 0.0631 & 0.0729 \\
Posterior & 0.0750 & 0.0714 & 0.0882 \\
\hline
\end{tabular}
\end{table}

\begin{table}[H]
\centering
\caption{The projected ground reaction forces on the socket \cite{winter2009biomechanics}.}
\label{Table 8}
\renewcommand{\arraystretch}{1.3}
\begin{tabular}{c c c}
\hline
\textbf{Phase/Force direction} & $F_y$ \textbf{[N]} & $F_z$ \textbf{[N]} \\
\hline
Heel strike & 120 & 580 \\
Mid-stance  & 92  & 582 \\
Push-off    & 112 & 578 \\
\hline
\end{tabular}
\end{table}

\begin{figure}[H]
    \centering
    \includegraphics[width=\linewidth]{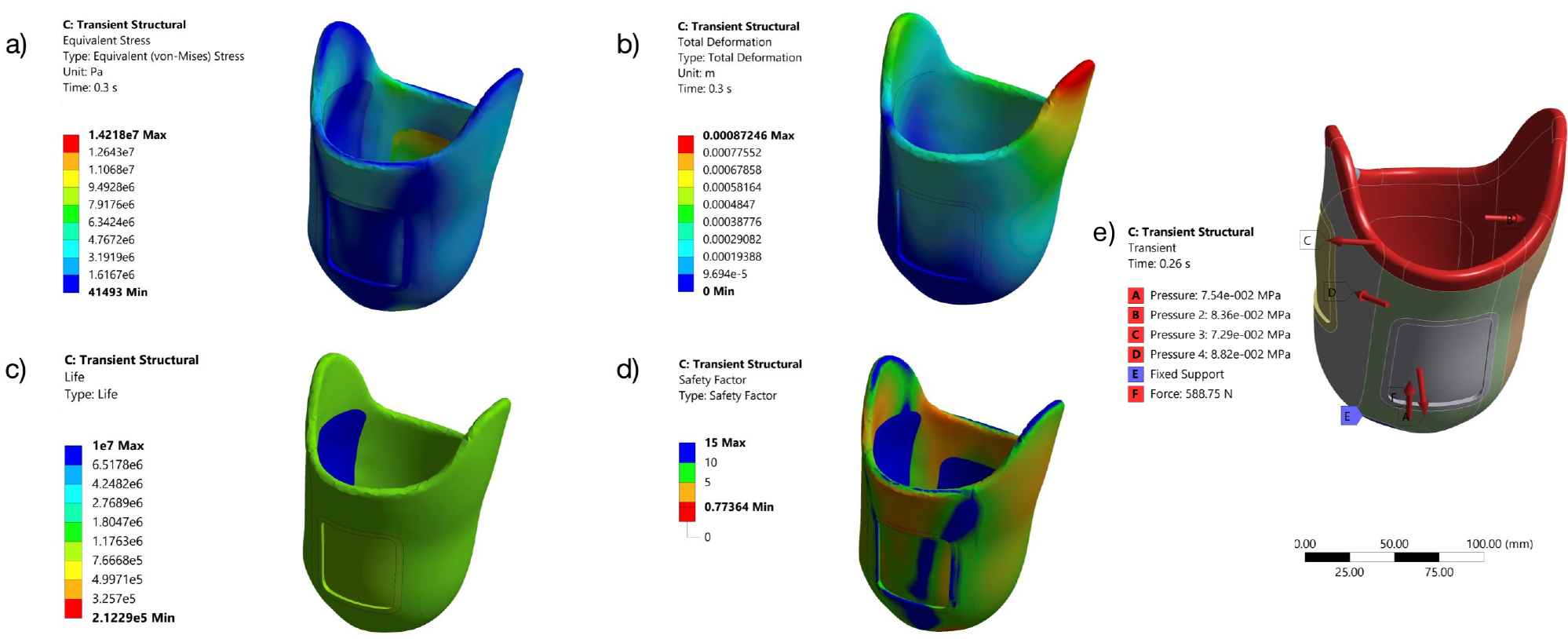}
    \caption{The Von-Mises results for the transient structural analysis (a), total deformation (b), life cycle (c), FOS (d), and Boundary conditions of the considered loads (e).}
    \label{stressanalysis}
\end{figure}

The Finite Element Analysis results are presented in Fig.\ref{stressanalysis}. The maximum stress is 14.218 MPa, and the peak deformation is 0.872 mm. The analysis indicates high stress at the edges and cutout edges, and minimal deformation. Moreover, the life cycle and factor of safety (FOS) are also calculated, as presented in Fig.\ref{stressanalysis}. The S-N curve values for Kevlar, TPU, PLA, and Carbon fiber materials are obtained from the literature \cite{baumann2021experimental, wang2019correlation, ezeh2019fatigue, burhan2018sn}. The analysis shows that the socket demonstrates a FOS ranging between 2.3 and 4.7 across different regions, while the predicted minimum fatigue life is \(2.12 \times 10^{6}\). However, in most areas the life cycle is around \(1.00 \times 10^{6}\). 

\subsection{Prosthetic Socket Manufacturing and Parts Connection}
\begin{figure}[H]
    \centering
    \includegraphics[width=\linewidth]{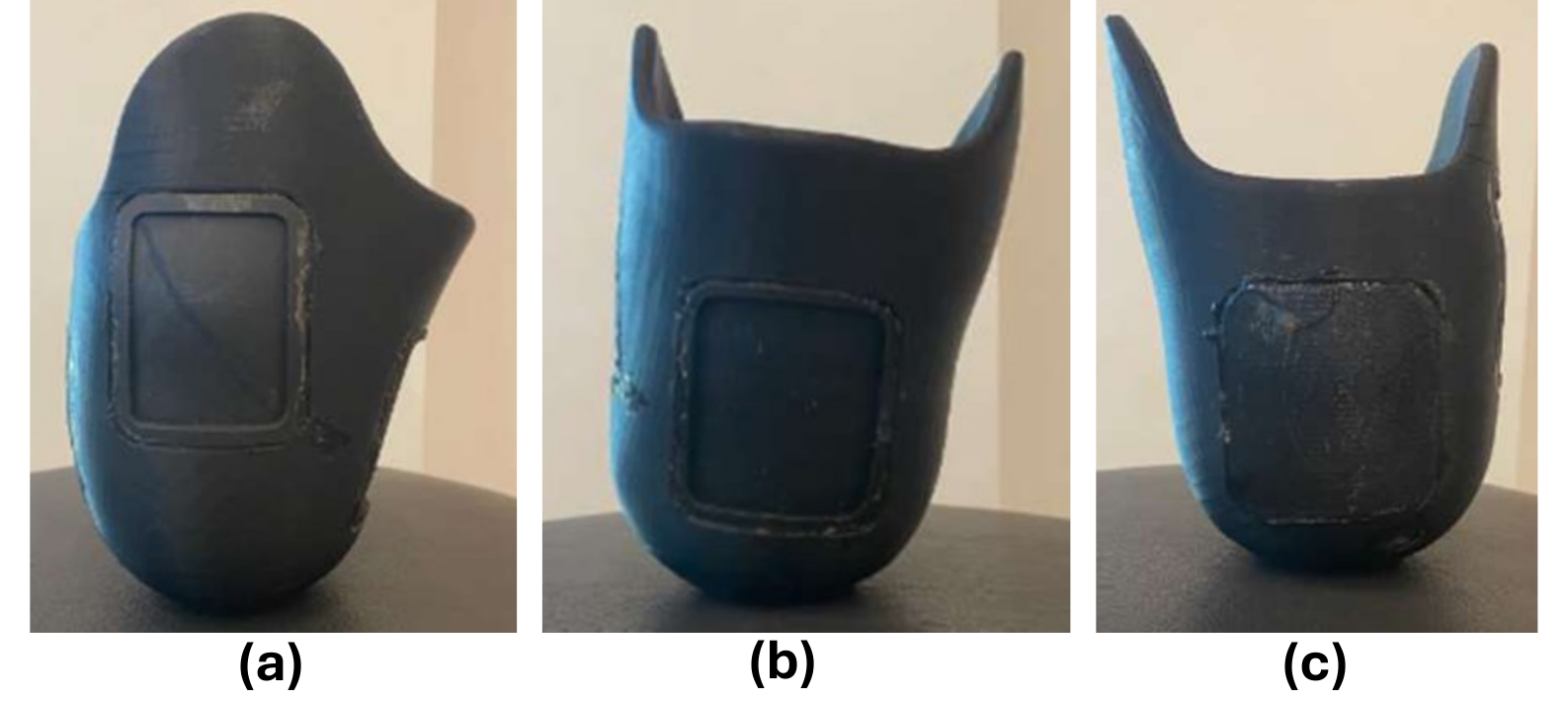}
    \caption{The prototyped i-Socket: Lateral view (Fibula) (a), Anterior view (Tibia) (b), Posterior view (Calf) (c).}
    \label{prot-socket}
\end{figure}

According to the PPT test results, different materials are selected for each area, and these parts are printed separately. The parts manufactured from tough PLA and TPU are printed using the Ultimaker (2Extended, Netherlands) 3D printer, and the parts from Carbon Fiber and Kevlar are prototyped with the composite 3D printer (Markforged, Industrial, x7, USA). In addition, the material’s printing direction and the type of stress applied should be considered because they impact the material's deforming point \cite{plesec2023numerical}. For the customized socket, the printing direction of the tough PLA part is upward (from the socket bottom to the top), and the printing direction for the TPU part is vertical (from the left side to the right side, the opposite of the tough PLA printing direction). Moreover, the printing direction of the carbon fiber parts is similar to the TPU part. Due to some printing errors, the outer surface of the Kevlar part will be the printing starting point. To connect these parts, glue is commonly used in prosthetics and sockets, such as the connection between the socket adapter and the pylon. This glue contains two materials, a siegelharz (Liquid form) and a mixture of dicyclohexyl phthalate and benzoyl peroxide (Powder form). The strength of the glue is tested during the static fatigue test. The prototyped socket is shown in Fig. \ref{prot-socket}. 
The mass of the participant's socket is equal to 0.36 kg, and the prototyped i-Socket mass is 0.28 kg. The composite structure of the socket is shown in Table \ref{Table 9}.

\begin{table}[H]
    \centering
    \caption{3D Printing Properties for Each Material}
    \label{Table 9}
    \renewcommand{\arraystretch}{1.3}
    \begin{tabular}{c c c c}
        \hline
        \textbf{Material} & \textbf{Infill \%} & \textbf{Infill pattern} & \textbf{Nozzle size [mm]} \\
        \hline
        TPU             & 45\% & Lines       & 0.6 \\
        Tough PLA       & 40\% & Tri-Hexagon & 0.6 \\
        Carbon fiber    & 27\% & Hexagonal   & 0.4 \\
        Kevlar          & 37\% & Triangular  & 0.4 \\
        \hline
    \end{tabular}
\end{table}

\section{Results}
To evaluate the pressure distribution inside the i-Socket and the kinematics of walking during the experiments, a 32-year-old male participant with a height of 165 cm and a weight of 60 kg, using an Ossur Vari-Flex prosthetic foot, participated. As stated earlier in the PPT experiments, the functional evaluations of the i-Socket have also been approved by the Clinical Research Ethics Committee of Istanbul Medipol University under document number E-10840098-772.02-7318. The participant in the functional evaluations confirmed his participation by signing the voluntary consent form.

\subsection{Mechanical Evaluation}
\begin{figure}[H]
    \centering
    \includegraphics[width=0.5\columnwidth]{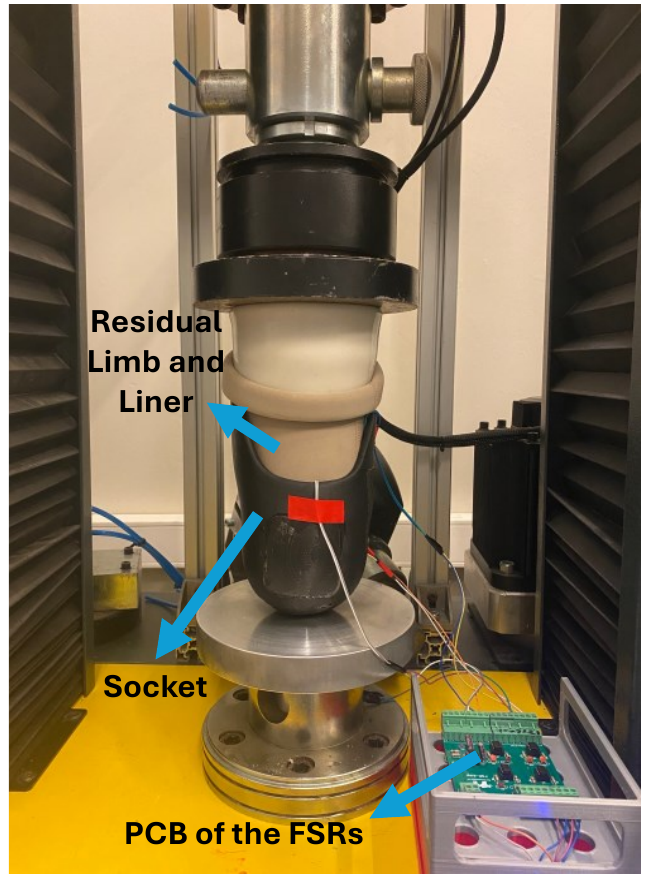}
    \caption{Mechanical evaluation test setup.}
    \label{instrontest}
\end{figure}

In order to evaluate the pressure disturbance on the desired regions in the socket under static loads, a mechanical test is conducted. The aim is to evaluate the socket strength and obtain a pressure distribution map in static analysis. The test setup includes an Instron machine (RAAGEN, ETM-300-S) that applies the force to the residual limb doll made of TPU material covered with a liner and three sensors placed at the Tibia, Fibula, and Calf areas. The sensor used in this test is the FlexiForce A502 Sensor (Tekscan Inc., South Boston, MA, USA), and the setup is shown in Fig.\ref{instrontest}. For a transtibial amputee, the prosthetic limb supports about 45\% of the total body weight \cite{fontes2021bodyweight}, which corresponds to 27 kg. For the static analysis, considering a safety factor of 1.5, and with an estimated 27 kg mass supported by the amputated limb, a force of 398 N is applied to the socket. To ensure a precise output, the force applied to the sensors is concentrated on a specific area by placing a thin circular object on the sensor’s surface. Moreover, the calibration process for the sensors is performed just before the test using masses of 0 kg, 10.25 kg, 19.85 kg, and 30.10 kg. For each mass, a corresponding voltage is obtained to derive the linear calibration equation, enabling the force/pressure values applied in this test to be obtained. In addition, the linear calibration equation for one of the sensors is \(y = 104.44x + 3.0086\), where x is the voltage output and y is the calculated force.

\begin{figure}[H]
    \centering
    \includegraphics[width=0.7\linewidth]{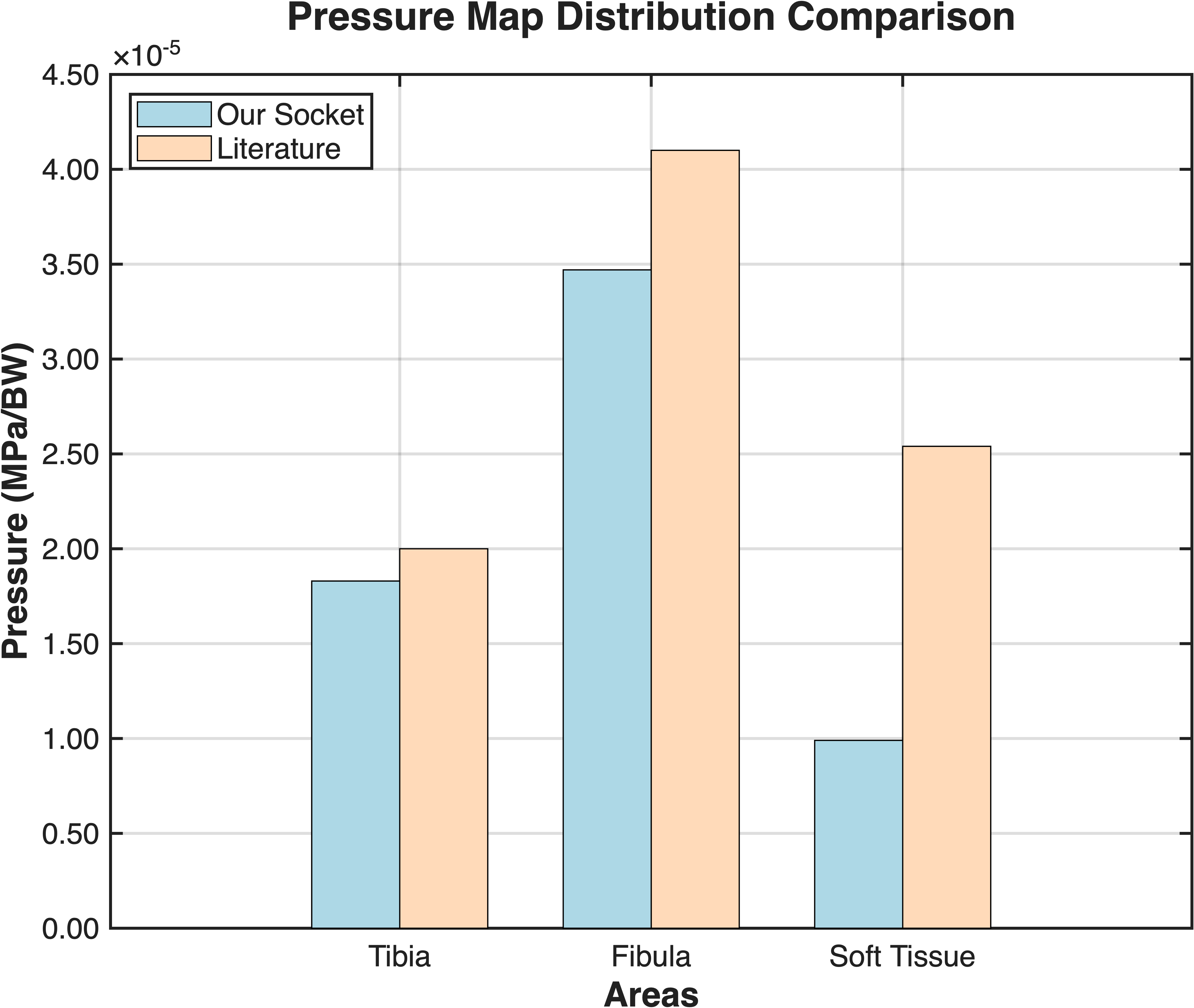}
    \caption{Comparison between the pressure values results during the mechanical evaluation and the pressure results in the literature.}
    \label{instrontestresults}
\end{figure}

The results for the three areas are presented in Fig.\ref{instrontestresults}, where the graphs show the raw pressure data and the filtered pressure data. Filtered data is obtained by using the data cleaner in MATLAB, and the cleaning parameters are the smoothing method set to moving mean, the smoothing parameter selected as the smoothing factor, and the smoothing factor chosen as 0.45. For a direct comparison with the results in the literature for a participant with a different mass, the pressure results for both obtained and the values in the literature are normalized according to the Body Weight (BW) of each participant. When normalized the results as BW and expressed them in MPa/BW, the calculated pressures for the proposed socket are Tibia = \(1.83 \times 10^{-5}\) MPa/BW, Fibula = \(3.47 \times 10^{-5}\) MPa/BW, and Calf  = \(0.99 \times 10^{-5}\) MPa/BW. In contrast, the pressure map distribution during static load in the literature for a 75 kg participant are \(2.00 \times 10^{-5}\) MPa/BW, \(4.10 \times 10^{-5}\) MPa/BW, and \(2.54 \times 10^{-5}\) MPa/BW for Tibia, Fibula, and Calf, respectively \cite{goh2003stump}.

\subsection{Pressure distribution Evaluation}
\begin{figure}[H]
    \centering
    \subfloat[]{
        \includegraphics[width=0.45\linewidth]{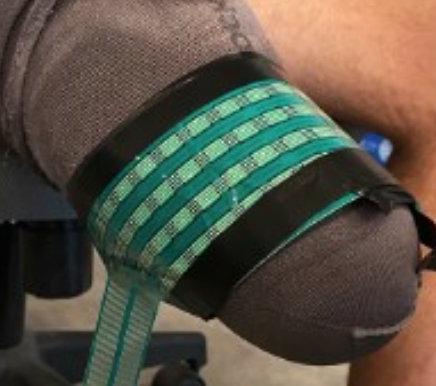}
        \label{Fsocketplacement}
    }\hfil
    \subfloat[]{
        \includegraphics[width=0.43\linewidth]{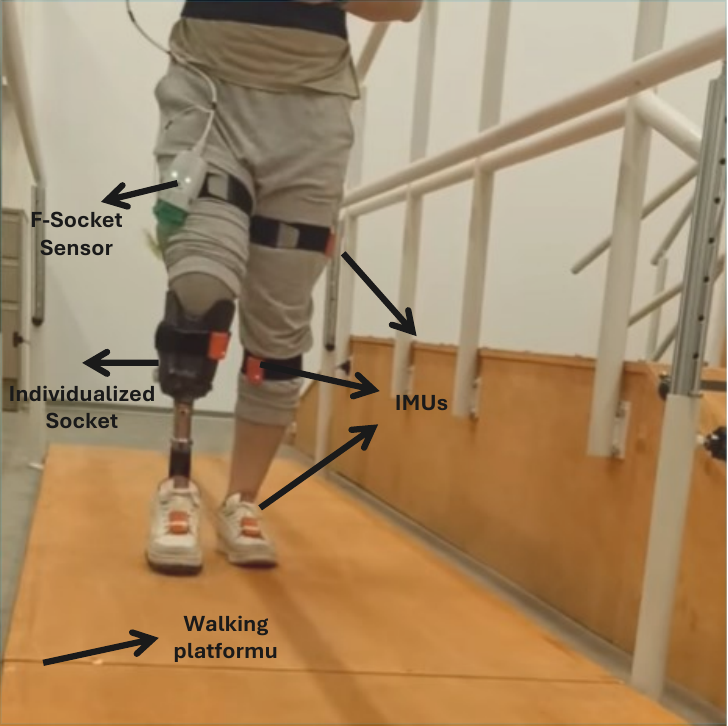}
        \label{fig:fsocketsetup}
    }
    \caption{The placement of the F-socket on the residual limb (a) and the test environment and the utilized sensors during the walking experiment (b).}
    \label{fig:fsocketcombined}
\end{figure}

To evaluate the pressure distribution inside the socket, the participant is instructed to walk over level ground at a self-selected, natural walking pace. During the test, pressure distribution is recorded using the F-Socket sensor, which is placed on the liner of the residual limb as shown in Fig.~\ref {Fsocketplacement}.

To calibrate the F-socket, known masses of 13.47 kg and 22.31 kg are applied to the sensor at specified frame intervals, and each mass is applied for 20 seconds \cite{peters2007signature}. All tests are conducted on the same day using both the participant’s own socket and the i-Socket. To protect the participant's prosthetic liner, TPU sheets with a thickness of 1 mm are printed and placed between the prosthetic liner and the socket. 

Due to the limited sensing area of the F-Socket sensor, simultaneous coverage of all residual limb interface regions is not feasible. Additionally, since the participant experienced less pain in the Calf region compared to the Tibia and Fibula regions, and because the Calf is considered a pressure-tolerant area, only the Tibia and Fibula areas are selected as concerned areas. As shown in Fig.~\ref{fig:fsocketsetup}, the F-Socket sensor is carefully positioned around the participant’s residual limb. The Tibia and Fibula regions are specifically marked both physically on the sensor surface and digitally within the analysis software to ensure consistent regional data tracking. 

\begin{figure*}[!t]
    \centering
    \subfloat[]{
        \includegraphics[width=0.8\columnwidth]{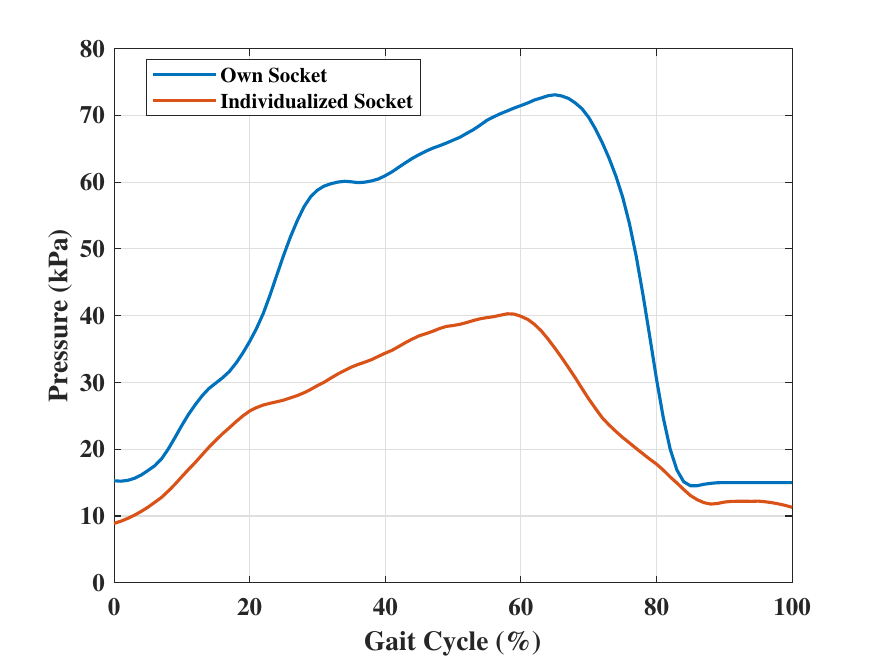}
        \label{fig:Tibiapressure}
    }\hfil
    \subfloat[]{
        \includegraphics[width=0.8\columnwidth]{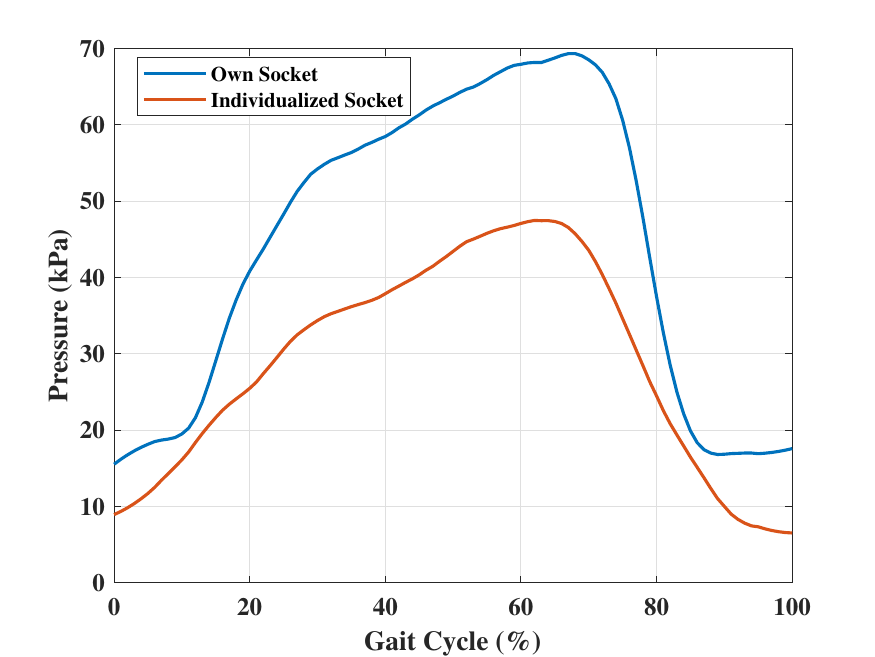}
        \label{fig:Fibulapressure}
    }
    \caption{Comparison of the pressure distribution results at the Tibia (a) and Fibula (b) regions during walking.}
    \label{fig:pressurecomparison}
\end{figure*}

The participant is asked to walk at a self-selected natural pace along a 7-meter walkway. During this trial, the average pressure distribution across a complete gait cycle is recorded for the Tibia and Fibula regions. The corresponding results are presented in Fig.~\ref{fig:Tibiapressure} and Fig.~\ref{fig:Fibulapressure}, respectively.

According to the results, the i-Socket reduced the peak pressure in the Tibia and Fibula regions by approximately 51\% and 50\%, respectively, throughout the gait cycle. Specifically, in the Tibia region, the maximum pressure measured with the i-Socket is 40.00~kPa at 55\% of the gait cycle, whereas the POS resulted in a peak pressure of 73.00~kPa at 64\%. Similarly, in the Fibula region, the i-Socket produced a maximum pressure of 47.97~kPa at 63\% of the gait cycle, compared to 69.92~kPa at 68\% when using the participant’s own socket. These findings indicate a substantial reduction in localized pressure due to the customized material and thickness distribution in the i-Socket design.

\subsection{Kinematic Evaluations}
\begin{figure}[t]
    \centering
    \includegraphics[width=0.5\columnwidth]{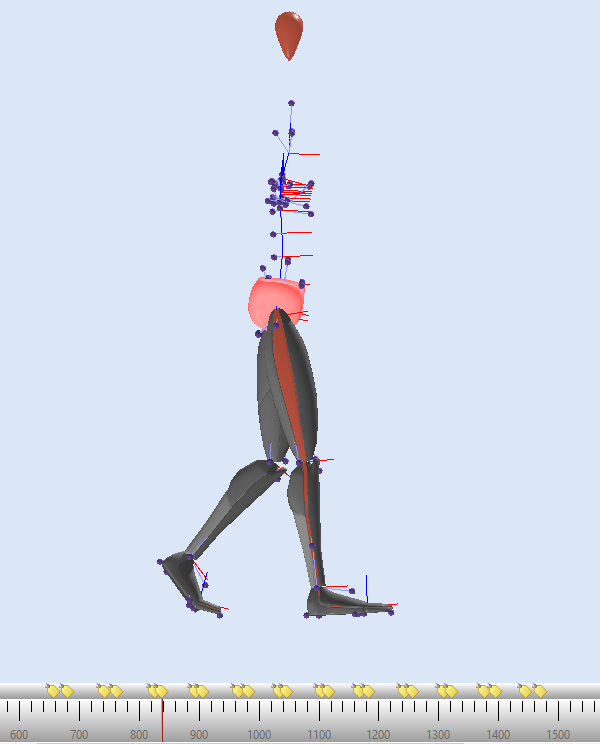}
    \caption{Identification of heel strike and toe-off instances during the gait cycles from the XSENS MVN analyzer software.}
    \label{fig:heelstrikelabeling}
\end{figure}

In the tests conducted with the transtibial amputee participant, inertial measurement units (IMUs) (XSENS MVN, Netherlands) sensors are utilized and placed on various segments of the body. A total of seven IMUs are positioned on the pelvis, right and left thighs, shanks, and feet. Prior to data collection, both the individual and relative positions of the sensors are calibrated using the XSENS MVN Analyze software.
Data are recorded at a sampling frequency of 100~Hz, with acquisition initiated at the beginning of each walking trial. To enhance the accuracy of joint angle analysis, gait cycles are identified based on heel strike and toe-off events, which are labeled within the XSENS MVN software. This labeling process is illustrated in Fig.~\ref{fig:heelstrikelabeling}. The walking experiments are conducted on a walkway and repeated three times for both the POS and the i-Socket. The test environment and sensor placements are illustrated in Fig.~\ref{fig:fsocketsetup}.

\subsubsection*{Join angles analyses}
\begin{figure*}[t]
    \centering
    \subfloat[]{
        \includegraphics[width=0.95\columnwidth]{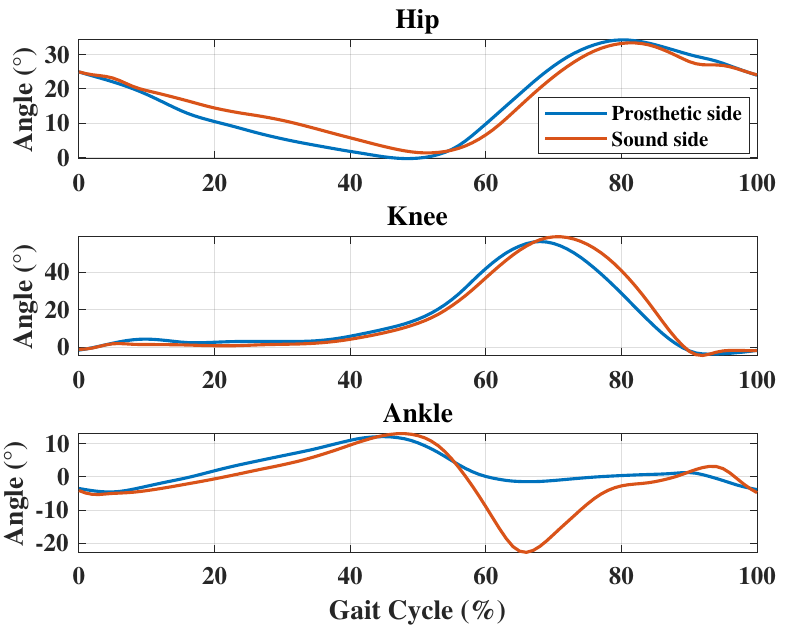}
        \label{fig:jointanglescustom}
    }
    \hfil 
    \subfloat[]{
        \includegraphics[width=0.95\columnwidth]{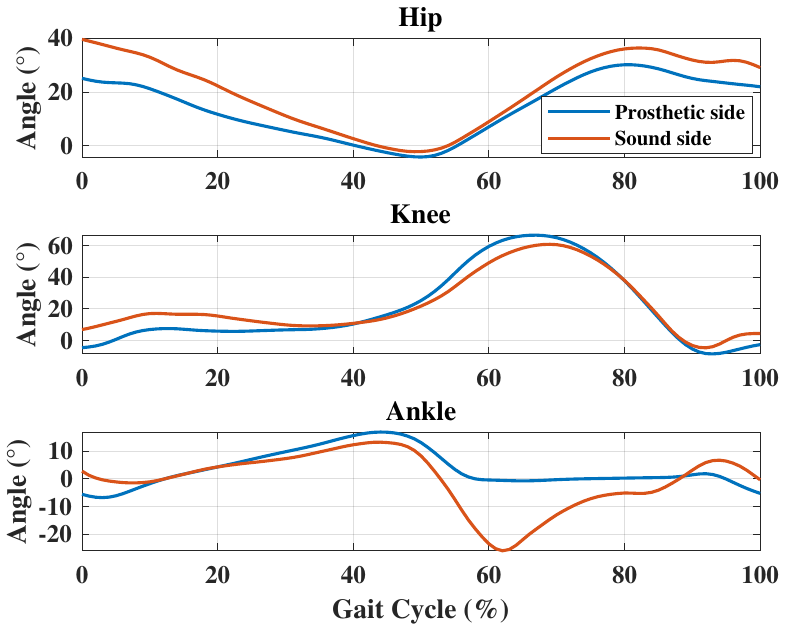}
        \label{fig:jointanglesconventional}
    }
    
    \caption{The joint angles of the lower limb joints while using the i-Socket (a) and the POS (b).}
    \label{fig:jointanglescomparison} 
\end{figure*}

For each walking cycle, the average joint angle values are computed over a complete gait cycle. These results are presented in Fig.~\ref{fig:jointanglescustom} and Fig.\ref{fig:jointanglesconventional}. Fig.~\ref{fig:jointanglescustom} and Fig.\ref{fig:jointanglesconventional} specifically compare the sagittal plane average angles of the hip, knee, and ankle joints for both the Prosthetic Side (PS) and Sound Side (SS) while the participant is using the individualized and his own socket, respectively.

Based on the results obtained from this figure, Table~\ref{tab:socketcomparison} presents the range of motion (RoM) of the joint angles and Pearson correlation coefficient values for the PS and SS across the analyzed joints for both individualized and POS. Since socket functionality is more dominant during the standing phase than during the swing phase, this table presents the results for the standing phase of the gait cycle.

\begin{table}[H]
    \centering
    \caption{Comparison of Range of Motion (RoM) and Correlation between i-Socket and POS across joints.}
    \label{tab:socketcomparison}
    \renewcommand{\arraystretch}{1.3} 
   \begin{tabular}{l l c c c c}
        \hline
        \multirow{2}{*}{\textbf{Joint}} & \multirow{2}{*}{\textbf{Side}} & \multicolumn{2}{c}{\textbf{RoM ($^\circ$)}} & \multicolumn{2}{c}{\textbf{Correlation (\%)}} \\
        \cline{3-6} 
        
         & & \textbf{i-Socket} & \textbf{POS} & \textbf{i-Socket} & \textbf{POS} \\
        \hline
        
        \multirow{2}{*}{Hip} & PS & 34.49 & 34.28 & \multirow{2}{*}{99} & \multirow{2}{*}{99.67} \\
         & SS & 31.95 & 41.73 & & \\
        \hline

        \multirow{2}{*}{Knee} & PS & 60.1 & 74.92 & \multirow{2}{*}{94.43} & \multirow{2}{*}{28.85} \\
         & SS & 63.39 & 65.31 & & \\
        \hline
        
        \multirow{2}{*}{Ankle} & PS & 16.67 & 23.64 & \multirow{2}{*}{98.83} & \multirow{2}{*}{96.45} \\
         & SS & 35.82 & 39.16 & & \\
        \hline
    \end{tabular}
\end{table}

\subsubsection*{Step time anlayses}
The stance phase duration of the prosthetic and sound limbs is a key indicator of the user's confidence in the socket and the amount of load transferred to the prosthetic limb. Therefore, during trials conducted at a normal walking speed, gait cycles are segmented by labeling heel strike and toe-off events as illustrated in Fig.\ref{fig:heelstrikelabeling}.

The results showed that the average stance phase duration is 63\% and 58.6\% of the gait cycle for the PS and the SS, respectively. This indicates that when using the i-Socket, the participant spent approximately 7.5\% less time in stance on the prosthetic limb, suggesting reduced loading and possibly lower trust in that limb.
\subsubsection*{Center of Mass Velocity}
\begin{figure}[t]
    \centering
 
    \includegraphics[width=\linewidth]{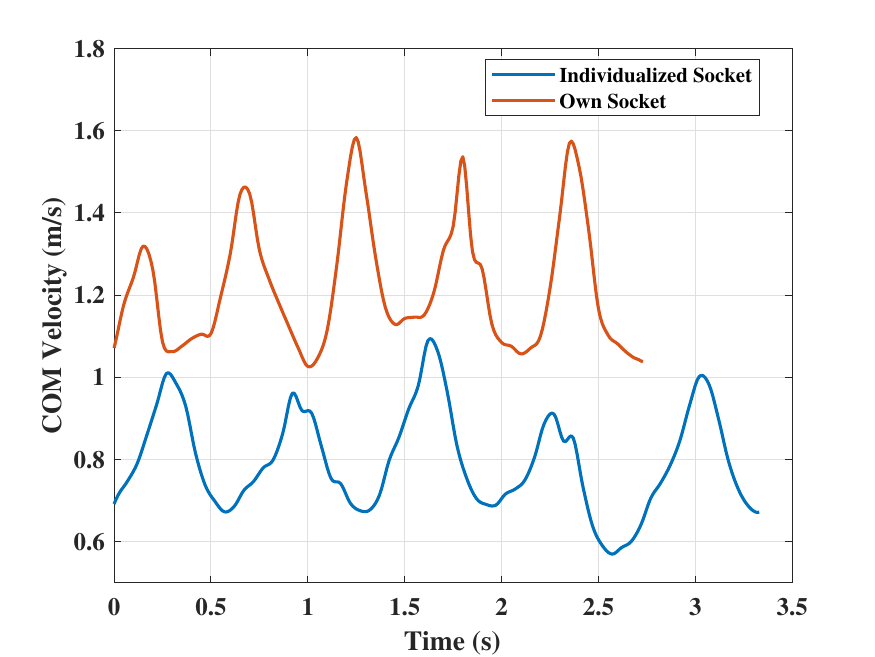}
    \caption{Forward CoM velocity during walking trials with the individualized and participant's own socket.}
    \label{fig:comvelocity}
\end{figure}

During the walking tests performed with both the i-Socket and the POS, the velocity of the body’s center of mass is recorded using the IMU sensors integrated into the XSENS system. Fig.~\ref{fig:comvelocity} illustrates the changes in Pelvis forward velocity over a 3.3 s data segment.

The analyses revealed that the average of the maximum velocity of the pelvis forward velocities is 1.39~m/s during walking with the POS, whereas this value decreased to 0.97~m/s when using the i-Socket. According to the kinematic analysis, the participant exhibited a greater range of motion in the knee joint while using his own socket, which contributed to the increased CoM velocity. In contrast, walking with the i-Socket resulted in reduced joint excursions at both the knee and ankle, leading to a lower CoM velocity.

\section{Discussion}
In this study, we aim to design and develop an individualized prosthetic socket using a multi-material/thickness concept to reduce the pressure between the residual limb and the socket for the transtibial amputees.
To reach this goal, first, pressure-sensitive/tolerant areas are specified based on the recommendations in literature \cite{physiopedia_lower_limb_sockets}. 
The selected conceptual design, which incorporates multi-material/thickness variations in different regions of the socket, is most compatible with the goals of our study, allowing for the control of both rigidity and flexibility of the socket. 
To finalize the material and thickness properties for each area of the prosthetic socket for a specific participant, the PPT test is conducted using different materials and thicknesses through a customized device that applies a perpendicular load to these regions. 
The results of structural analysis based on the projected ground reaction forces on the socket show that the Fibula and Tibia, as the pressure-sensitive areas, have the minimum stresses, and the socket is strong enough against the applied forces. This finding supports the idea that an i-Socket can decrease the high pressure inside the socket during the gait cycle. The final design of the i-Socket is manufactured, and the mass of it is 0.28 kg, which is 22\% lighter than the POS and 5 times lighter than a similar multi-material socket \cite{sengeh2013variable}. This lightweight prototype not only ensures the safety of the interface between the residual limb and the prosthesis for the participant, but also can increase the ease during walking \cite{gariboldi2022structural}. 
To ensure the socket's strength against loads, a mechanical evaluation test is conducted to assess the pressure map distribution of the prototyped socket. This evaluation is obtained using a universal compression machine with 45\% of the applied mass on the prosthetic socket \cite{fontes2021bodyweight}. The results show that the normalized pressure map distribution for a participant with a mass of 60 kg reduces by 8.5\% for the tibia, 15.4\% for the Fibula, and 61.0\% for the Calf compared to the same evaluation in the literature \cite{goh2003stump}.
According to the pressure analysis in the functional evaluations with the transtibial amputee, the pressure values decreased by 45\% and 31\%, respectively, in both the Tibia and Fibula regions, when wearing the i-Socket compared to the POS. Considering the effects of the CoM velocity during the experiments as the normalized pressure, the decrease in pressure values in the Tibia and Fibula regions is 21\% and 2\%, respectively. Compared to the multi-material socket \cite{sengeh2013variable}, the reduction in the Tibia region of the i-Socket is 168\% more, although the self-selected speed is higher during walking with the i-Socket.
From a kinematic perspective, greater similarity in joint trajectories is observed between the prosthetic and sound limbs when using the i-Socket. Particularly in the correlation coefficient value between the Ps and SS in the knee and ankle joint during the stance phase of the gait cycle (65\% and 2\% more symmetry in the knee and ankle joints, respectively). This suggests improved interlimb synchronization and enhanced gait symmetry when using the i-Socket. Additionally, the self-selected speed during walking with the i-Socket is 15\% more than the same metric in the multi-material socket \cite{sengeh2013variable}.
The differences in CoM velocity between using the i-Socket and POS may be attributed to the participant’s familiarity with his own socket, higher perceived trust in it, and a lack of full adaptation to the conceptual differences with the i-Socket. Furthermore, during initial trials with the new socket, a cautious gait pattern may have emerged, limiting step length and joint swing, thereby reducing CoM velocity. The findings suggest a direct relationship between CoM velocity and joint range of motion, which appears to be closely linked to the user’s adaptation process to the socket. Overall, these findings suggest that the i-Socket not only has the capability to reduce the pressure distribution in the concerned regions but also can have a positive influence on gait symmetry and speed.

In this study, we only focus on three main regions (Tibia, Fibula, and Calf) due to their importance for the comfort and stability of the participant during the activities of daily living (ADL). In future works, the additional areas will be considered for specified participant based on their PPT results in those regions to determine different material/thickness properties. Additionally, this manuscript includes only PLA material for the non-pressure-sensitive/tolerant areas based on its biocompatibility and high strength properties \cite{lestari2024optimization}. A further material selection procedure for these areas will be considered in future works. Since, in the design stage of this study, the PPT and applied pressure values for the amputee participant were not available, the values in the literature are used for the concerned areas. Because of the limitations of the compression machine in the mechanical evaluation, only the static loads are considered to investigate the load-bearing and pressure distribution behavior of the different regions of the socket. For the dynamic loads of different ADL and fatigue analyses, experiments are also planned to be conducted with proper compression machines in future work. Additionally, it is planned to evaluate the functionality and the pressure distribution reducing capability of the i-Socket with real users in future work during different ADL.

\section{Conclusion}
In this study, the design, development, and functional evaluation of an individualized transtibial prosthetic socket that includes multi-material/ thickness regions is proposed. Using the structural analyses and simulations for different conceptual designs, the final design is selected based on the requirements of the study, such as comfort and less pressure in the concerned areas. The material and thickness of these regions are determined based on the results of the PPT tests with the transtibial amputee participant. The finalized design of the socket is manufactured, and the mass of it is 22\% less than the POS mass, which can be considered as an important parameter in using the socket. 
Based on the results from the mechanical evaluation for a 60 kg transtibial amputee participant, the socket pressure decreased by 8.5\% in the Tibia, 15.4\% in the Fibula, and 61.0\% in the Calf regions compared with values reported in the literature. The results of the functional evaluations show that 45\% and 31\% reduction in the pressure of the Tibia and Fibula regions, respectively. Additionally, the self-selected walking speed is increased by 15\% compared to the similar studies in the literature.
For future studies, to obtain the optimal design for each participant, the mass and strength of the socket will be optimized based on the different parameters of designing and prototyping. Additionally, to evaluate the socket during the ADL and investigate the pressure inside the socket for these activities, the experiments with the transtibial amputee will be performed, and the results will be compared with the POS.

\section*{Acknowledgment}
This study was partially supported by the Scientific and Technological Research Council of Turkey (TUBITAK) under Grant Number 123M351. The authors thank TUBITAK for their support. The authors would like to extend their appreciation to Ulas Yildirim, Canay Isil, Farhan Raza, and Prof. Dr. Guven Yapici for their assistance with mechanical testing.




%
\bibliographystyle{ieeetr}
\bibliography{mybib}

\begin{thebibliography}{10}

\bibitem{who2017}
{World Health Organization} and USAID, {\em {WHO standards for prosthetics and
  orthotics}}.
\newblock Geneva: World Health Organization, 2017.

\bibitem{varsavas2022review}
S.~D. Varsavas, F.~Riemelmoser, F.~Arbeiter, and L.-M. Faller, ``A review of
  parameters affecting success of lower-limb prosthetic socket and liners and
  implementation of 3d printing technologies,'' {\em Materials Today:
  Proceedings}, vol.~70, pp.~425--430, 2022.

\bibitem{noll2017physically}
V.~Noll, N.~Eschner, C.~Schumacher, P.~Beckerle, and S.~Rinderknecht, ``A
  physically-motivated model describing the dynamic interactions between
  residual limb and socket in lower limb prostheses,'' {\em Current Directions
  in Biomedical Engineering}, vol.~3, no.~1, pp.~15--18, 2017.

\bibitem{xiaohong2004dynamic}
J.~Xiaohong, Z.~Ming, W.~Rencheng, and J.~Dewen, ``Dynamic investigation of
  interface stress on below-knee residual limb in a prosthetic socket,'' {\em
  Tsinghua Science and Technology}, vol.~9, no.~6, pp.~680--683, 2004.

\bibitem{price2019design}
M.~A. Price, P.~Beckerle, and F.~C. Sup, ``Design optimization in lower limb
  prostheses: A review,'' {\em IEEE Transactions on Neural Systems and
  Rehabilitation Engineering}, vol.~27, no.~8, pp.~1574--1588, 2019.

\bibitem{dumbleton2009dynamic}
T.~Dumbleton, A.~W. Buis, A.~McFadyen, B.~F. McHugh, G.~McKay, K.~D. Murray,
  and S.~Sexton, ``Dynamic interface pressure distributions of two transtibial
  prosthetic socket concepts.,'' {\em Journal of Rehabilitation Research \&
  Development}, vol.~46, no.~3, 2009.

\bibitem{ibarra2020interface}
S.~Ibarra~Aguila, G.~J. S{\'a}nchez, E.~E. Sauvain, B.~Alemon, R.~Q.
  Fuentes-Aguilar, and J.~C. Huegel, ``Interface pressure system to compare the
  functional performance of prosthetic sockets during the gait in people with
  trans-tibial amputation,'' {\em Sensors}, vol.~20, no.~24, p.~7043, 2020.

\bibitem{paterno2018sockets}
L.~Patern{\`o}, M.~Ibrahimi, E.~Gruppioni, A.~Menciassi, and L.~Ricotti,
  ``Sockets for limb prostheses: a review of existing technologies and open
  challenges,'' {\em IEEE Transactions on Biomedical Engineering}, vol.~65,
  no.~9, pp.~1996--2010, 2018.

\bibitem{gariboldi2022structural}
F.~Gariboldi, D.~Pasquarelli, and A.~G. Cutti, ``Structural testing of
  lower-limb prosthetic sockets: A systematic review,'' {\em Medical
  engineering \& physics}, vol.~99, p.~103742, 2022.

\bibitem{mak2001state}
A.~F. Mak, M.~Zhang, and D.~A. Boone, ``State-of-the-art research in lower-limb
  prosthetic biomechanics-socket interface: a review.,'' {\em Journal of
  Rehabilitation Research \& Development}, vol.~38, no.~2, 2001.

\bibitem{yiugiter2002comparison}
K.~Yi{\u{g}}iter, G.~{\c{S}}ener, and K.~Bayar, ``Comparison of the effects of
  patellar tendon bearing and total surface bearing sockets on prosthetic
  fitting and rehabilitation,'' {\em Prosthetics and orthotics international},
  vol.~26, no.~3, pp.~206--212, 2002.

\bibitem{bagheripour2022design}
B.~Bagheripour, M.~A. Mardani, B.~Hajiaghaei, A.~Biglarian, T.~Babaee, and
  H.~Pezham, ``Design and evaluation of a prosthetic socket for a patient with
  diabetic-related transtibial amputation: A case report,'' {\em Clinical Case
  Reports}, vol.~10, no.~9, p.~e6276, 2022.

\bibitem{ramlee2024investigation}
M.~H. Ramlee, M.~I. Ammarullah, N.~S. Mohd~Sukri, N.~S. Faidzul~Hassan, M.~H.
  Baharuddin, and M.~R. Abdul~Kadir, ``Investigation on three-dimensional
  printed prosthetics leg sockets coated with different reinforcement
  materials: analysis on mechanical strength and microstructural,'' {\em
  Scientific Reports}, vol.~14, no.~1, p.~6842, 2024.

\bibitem{faustini2006experimental}
M.~C. Faustini, R.~R. Neptune, R.~H. Crawford, W.~E. Rogers, and G.~Bosker,
  ``An experimental and theoretical framework for manufacturing prosthetic
  sockets for transtibial amputees,'' {\em IEEE Transactions on neural systems
  and rehabilitation engineering}, vol.~14, no.~3, pp.~304--310, 2006.

\bibitem{comotti2015multi}
C.~Comotti, D.~Regazzoni, C.~Rizzi, and A.~Vitali, ``Multi-material design and
  3d printing method of lower limb prosthetic sockets,'' in {\em Proceedings of
  the 3rd 2015 workshop on ICTs for improving patients rehabilitation research
  techniques}, pp.~42--45, 2015.

\bibitem{sengeh2013variable}
D.~M. Sengeh and H.~Herr, ``A variable-impedance prosthetic socket for a
  transtibial amputee designed from magnetic resonance imaging data,'' {\em
  JPO: Journal of Prosthetics and Orthotics}, vol.~25, no.~3, pp.~129--137,
  2013.

\bibitem{physiopedia_lower_limb_sockets}
{Physiopedia contributors}, ``Lower limb prosthetic sockets and suspension
  systems.''
  \url{https://www.physio-pedia.com/Lower_Limb_Prosthetic_Sockets_and_Suspension_Systems},
  July 2022.
\newblock Accessed: 2022-07-11.

\bibitem{novacheck1998biomechanics}
T.~F. Novacheck, ``The biomechanics of running,'' {\em Gait \& Posture},
  vol.~7, pp.~77--95, 1998.

\bibitem{convery1999socket}
P.~Convery and A.~Buis, ``Socket/stump interface dynamic pressure distributions
  recorded during the prosthetic stance phase of gait of a trans-tibial amputee
  wearing a hydrocast socket,'' {\em Prosthetics and Orthotics International},
  vol.~23, no.~2, pp.~107--112, 1999.

\bibitem{lee2005regional}
W.~C. Lee, M.~Zhang, and A.~F. Mak, ``Regional differences in pain threshold
  and tolerance of the transtibial residual limb: including the effects of age
  and interface material,'' {\em Archives of physical medicine and
  rehabilitation}, vol.~86, no.~4, pp.~641--649, 2005.

\bibitem{rolke2005deep}
R.~Rolke, K.~A. Campbell, W.~Magerl, and R.-D. Treede, ``Deep pain thresholds
  in the distal limbs of healthy human subjects,'' {\em European Journal of
  Pain}, vol.~9, no.~1, pp.~39--48, 2005.

\bibitem{polianskis2001computer}
R.~Polianskis, T.~Graven-Nielsen, and L.~Arendt-Nielsen, ``Computer-controlled
  pneumatic pressure algometry—a new technique for quantitative sensory
  testing,'' {\em European journal of pain}, vol.~5, no.~3, pp.~267--277, 2001.

\bibitem{lee2019evaluation}
H.~Lee, R.-i. Eom, and Y.~Lee, ``Evaluation of the mechanical properties of
  porous thermoplastic polyurethane obtained by 3d printing for protective
  gear,'' {\em Advances in Materials Science and Engineering}, vol.~2019,
  no.~1, p.~5838361, 2019.

\bibitem{markforged_composites_datasheet}
{Markforged}, ``Composites data sheet.''
  \url{https://static.markforged.com/downloads/composites-data-sheet.pdf}, n.d.
\newblock Accessed: \today.

\bibitem{winter2009biomechanics}
D.~A. Winter, {\em Biomechanics and motor control of human movement}.
\newblock John wiley \& sons, 2009.

\bibitem{baumann2021experimental}
A.~Baumann and J.~Hausmann, ``Experimental investigation of instabilities on
  different scales in compressive fatigue testing of composites,'' {\em Journal
  of Composites Science}, vol.~5, no.~4, p.~114, 2021.

\bibitem{wang2019correlation}
C.~Wang, T.~Stiller, A.~Hausberger, G.~Pinter, F.~Gr{\"u}n, and T.~Schwarz,
  ``Correlation of tribological behavior and fatigue properties of filled and
  unfilled tpus,'' {\em Lubricants}, vol.~7, no.~7, p.~60, 2019.

\bibitem{ezeh2019fatigue}
O.~Ezeh and L.~Susmel, ``Fatigue strength of additively manufactured
  polylactide (pla): effect of raster angle and non-zero mean stresses,'' {\em
  International Journal of Fatigue}, vol.~126, pp.~319--326, 2019.

\bibitem{burhan2018sn}
I.~Burhan and H.~S. Kim, ``Sn curve models for composite materials
  characterisation: an evaluative review,'' {\em Journal of Composites
  Science}, vol.~2, no.~3, p.~38, 2018.

\bibitem{plesec2023numerical}
V.~Plesec, J.~Humar, P.~Dobnik-Dubrovski, and G.~Harih, ``Numerical analysis of
  a transtibial prosthesis socket using 3d-printed bio-based pla,'' {\em
  Materials}, vol.~16, no.~5, p.~1985, 2023.

\bibitem{fontes2021bodyweight}
C.~H. d.~S. Fontes~Filho, C.~T. Laett, U.~F. Gavil{\~a}o, J.~C.~d. Campos~Jr,
  D.~J. d.~A. Alexandre, V.~R. Cossich, and E.~B.~d. Sousa, ``Bodyweight
  distribution between limbs, muscle strength, and proprioception in traumatic
  transtibial amputees: a cross-sectional study,'' {\em Clinics}, vol.~76,
  p.~e2486, 2021.

\bibitem{goh2003stump}
J.~Goh, P.~Lee, and S.~Chong, ``Stump/socket pressure profiles of the pressure
  cast prosthetic socket,'' {\em Clinical Biomechanics}, vol.~18, no.~3,
  pp.~237--243, 2003.

\bibitem{peters2007signature}
J.~Peters, S.~Howington, J.~Ballard, and L.~Lynch, ``Signature evaluation for
  thermal infrared countermine and ied detection systems,'' in {\em 2007 DoD
  High Performance Computing Modernization Program Users Group Conference},
  pp.~238--246, IEEE, 2007.

\bibitem{lestari2024optimization}
W.~D. Lestari, N.~Adyono, A.~K. Faizin, A.~Haqiyah, K.~H. Sanjaya, A.~Nugroho,
  W.~Kusmasari, and M.~I. Ammarullah, ``Optimization of 3d printed parameters
  for socket prosthetic manufacturing using the taguchi method and response
  surface methodology,'' {\em Results in Engineering}, vol.~21, p.~101847,
  2024.

\end{thebibliography}

%




\end{document}